\documentclass[draft]{agujournal}
\draftfalse

\usepackage[none]{hyphenat}

\journalname{JGR-Planets}

\begin{document}

%% ------------------------------------------------------------------------ %%
%  Title
% 
% (A title should be specific, informative, and brief. Use
% abbreviations only if they are defined in the abstract. Titles that
% start with general keywords then specific terms are optimized in
% searches)
%
%% ------------------------------------------------------------------------ %%

% Example: \title{This is a test title}

\title{Effect of Re-impacting Debris on the Solidification of the Lunar Magma Ocean}

%% ------------------------------------------------------------------------ %%
%
%  AUTHORS AND AFFILIATIONS
%
%% ------------------------------------------------------------------------ %%

% Authors are individuals who have significantly contributed to the
% research and preparation of the article. Group authors are allowed, if
% each author in the group is separately identified in an appendix.)

% List authors by first name or initial followed by last name and
% separated by commas. Use \affil{} to number affiliations, and
% \thanks{} for author notes.  
% Additional author notes should be indicated with \thanks{} (for
% example, for current addresses). 

% Example: \authors{A. B. Author\affil{1}\thanks{Current address, Antartica}, B. C. Author\affil{2,3}, and D. E.
% Author\affil{3,4}\thanks{Also funded by Monsanto.}}

\authors{Viranga Perera\affil{1}\thanks{Current address: Applied Physics Lab, Johns Hopkins University, 11100 Johns Hopkins Road, Laurel, MD 20723, USA.}, Alan P. Jackson\affil{2,1}, Linda T. Elkins-Tanton\affil{1}, Erik Asphaug\affil{3,1}}

\affiliation{1}{School of Earth and Space Exploration, Arizona State University, P.O. Box 876004, Tempe, AZ 85287-6004, USA.}
\affiliation{2}{Centre for Planetary Sciences, University of Toronto, 1265 Military Trail, Toronto, ON, M1C 1A4, Canada.}
\affiliation{3}{Lunar and Planetary Laboratory, University of Arizona, P.O. Box 210092, Tucson, AZ 85721, USA.}
%(repeat as many times as is necessary)

%% Corresponding Author:
% Corresponding author mailing address and e-mail address:

% (include name and email addresses of the corresponding author.  More
% than one corresponding author is allowed in this LaTeX file and for
% publication; but only one corresponding author is allowed in our
% editorial system.)  

% Example: \correspondingauthor{First and Last Name}{email@address.edu}

\correspondingauthor{Viranga Perera}{viranga.perera@jhuapl.edu}

%% Keypoints, final entry on title page.

% Example: 
% \begin{keypoints}
% \item	List up to three key points (at least one is required)
% \item	Key Points summarize the main points and conclusions of the article
% \item	Each must be 100 characters or less with no special characters or punctuation 
% \end{keypoints}

%  List up to three key points (at least one is required)
%  Key Points summarize the main points and conclusions of the article
%  Each must be 100 characters or less with no special characters or punctuation 

\begin{keypoints}
\item At least a lunar mass of debris would have escaped the Earth-Moon system after the Moon forming impact
\item The subsequent return of that debris may have significantly affected the thermal evolution of the Moon
\item Re-impacting debris may have either expedited or delayed the solidification of the Lunar Magma Ocean
\end{keypoints}

%% ------------------------------------------------------------------------ %%
%
%  ABSTRACT
%
% A good abstract will begin with a short description of the problem
% being addressed, briefly describe the new data or analyses, then
% briefly states the main conclusion(s) and how they are supported and
% uncertainties. 
%% ------------------------------------------------------------------------ %%

%% \begin{abstract} starts the second page 

\begin{abstract}
Anorthosites that comprise the bulk of the lunar crust are believed to have formed during solidification of a Lunar Magma Ocean (LMO) in which these rocks would have floated to the surface. This early flotation crust would have formed a thermal blanket over the remaining LMO, prolonging solidification. Geochronology of lunar anorthosites indicates a long timescale of LMO cooling, or re-melting and re-crystallization in one or more late events. To better interpret this geochronology, we model LMO solidification in a scenario where the Moon is being continuously bombarded by returning projectiles released from the Moon-forming giant impact. More than one lunar mass of material escaped the Earth-Moon system onto heliocentric orbits following the giant impact, much of it to come back on returning orbits for a period of 100 Myr. If large enough, these projectiles would have punctured holes in the nascent floatation crust of the Moon, exposing the LMO to space and causing more rapid cooling. We model these scenarios using a thermal evolution model of the Moon that allows for production (by cratering) and evolution (solidification and infill) of holes in the flotation crust that insulates the LMO. For effective hole production, solidification of the magma ocean can be significantly expedited, decreasing the cooling time by more than a factor of 5. If hole production is inefficient, but shock conversion of projectile kinetic energy to thermal energy is efficient, then LMO solidification can be somewhat prolonged, lengthening the cooling time by 50\% or more.

\end{abstract}

%% ------------------------------------------------------------------------ %%
%
%  TEXT
%
%% ------------------------------------------------------------------------ %%

%%% Suggested section heads:
% \section{Introduction}
% 
% The main text should start with an introduction. Except for short
% manuscripts (such as comments and replies), the text should be divided
% into sections, each with its own heading. 

% Headings should be sentence fragments and do not begin with a
% lowercase letter or number. Examples of good headings are:

% \section{Materials and Methods}
% Here is text on Materials and Methods.
%
% \subsection{A descriptive heading about methods}
% More about Methods.
% 
% \section{Data} (Or section title might be a descriptive heading about data)
% 
% \section{Results} (Or section title might be a descriptive heading about the
% results)
% 
% \section{Conclusions}

% INTRODUCTION
\section{Introduction}

The Moon likely coalesced from debris in the aftermath of a giant impact between the proto-Earth and another planet-sized body \citep{Daly_1946, Hartmann_1975, Cameron_1976}. This Giant Impact Model explains the high angular momentum of the Earth--Moon system, the iron depletion of the Moon relative to the Earth, and the Moon's volatile depletion \citep{Wolf_1980, Taylor_2006, Taylor_2014}. After several iterations, a Canonical Giant Impact Model of a low-velocity, glancing impact by a Mars-sized body developed \citep[see][for a review]{Canup_2004}. In the Canonical model, the Moon is predominantly composed of material from the impactor; however, recent geochemical analyses show that the Earth and the Moon have nearly identical isotopic signatures \citep[e.g.][]{Touboul_2007, Spicuzza_2007, Zhang_2012}. As a result, several works have proposed modifications to the Giant Impact Model to account for the isotopic similarities \citep[e.g.][]{Pahlevan_2007, Canup_2012, Cuk_2012, Reufer_2012, MastrobuonoBattisti_2015, Cuk_2016, Rufu_2017}. Though the Giant Impact Model will undoubtedly continue to be revised and improved, it still is the accepted mechanism for the formation of the Moon \citep{Asphaug_2014, Barr_2016}.

\subsection{Initial Thermal State of the Moon}
\label{sec:Introduction:ThermalState}

The initial thermal state of a newly formed planet is primarily determined by how long it takes to form and how efficient accretionary impacts are at depositing thermal energy. Formation time is particularly important for two reasons. First, it will determine how much of gravitational potential energy is thermally radiated away and how much is used to heat constituent material. Second, it will determine if hot disk material will be accreted quickly (e.g. silicate material in the debris disk were likely between 2,500 to 5,000 K after the Giant Impact \citep{Canup_2004}). Thus, the Moon would have been initially molten if it accreted rapidly. In that case, the debris would have been hot and the Moon's gravitational binding energy, which, per unit mass, is comparable to the latent heat of silicates \citep{Pritchard_2000}, would have been used to melt constituent material. Though the Moon likely accreted rapidly, accretionary models vary in their estimates as to how long the Moon took to acquire the majority of its mass. For the Canonical model, that period is generally thought to be between a month to a year \citep{Ida_1997, Kokubo_2000, Takeda_2001}. Additionally, how efficient accretionary impacts are at depositing thermal energy is subject to considerable uncertainty. Past works have assumed that a certain fraction of the accretion energy was deposited into the planet as thermal energy \citep[e.g.][]{Kaula_1979, Ransford_1980, Squyres_1988, Senshu_2002, Merk_2006}; however, the temperature of the planet at the end of accretion is strongly dependent on those assumptions \citep{Stevenson_1986}. Given these uncertainties, while dynamics suggests that an early Lunar Magma Ocean (LMO) is likely, it is currently not possible to be definitive regarding the initial thermal state of the Moon from a purely dynamical perspective.

An alternative, yet complementary, approach to characterizing the initial thermal state of the Moon is by geochemical analyses of lunar samples. Early work on Apollo samples found ferroan anorthosite (FAN) rock fragments \citep{Wood_1970a}. From that observation it was inferred that the early lunar crust was made from anorthositic rocks that floated to the surface of a LMO \citep{Wood_1970b}. Anorthosite rocks were buoyant due to the low density of its primary mineral plagioclase feldspar. Recent reflectance spectral data are consistent with this scenario since they show the presence of pure anorthosite on a large fraction of the lunar surface \citep{Yamamoto_2012}. The europium (\textit{Eu}) anomaly is further evidence for a past LMO. \textit{Eu} is drawn to plagioclase feldspar and as such is enriched in the lunar crust and depleted in the mantle \citep{Philpotts_1970, Wakita_1970}. Additionally, incompatible KREEP elements (i.e. potassium [K], rare earth elements [REE], and phosphorous [P]) that exists on the lunar surface are likely from residual liquid of the LMO (i.e. ur-KREEP) \citep{Warren_1979}. Some works have questioned the existence of an LMO \citep[e.g.][]{Walker_1983, Longhi_1985, Longhi_2003, Boyet_2007}, but the amalgamation of evidence suggests that a LMO existed \citep[for a review see][]{ElkinsTanton_2012}.

To understand the thermal evolution of the Moon, it is important to estimate the initial depth of the LMO and the time that it took to solidify. The initial depth has been estimated by starting with an estimate for the lunar crustal thickness and arguing that a LMO of a certain initial composition (viz. $Al_{2}O_{3}$) needed to have been a particular depth to have produced that crust by fractional crystallization \citep{Warren_1985, Yamamoto_2012}. For that depth estimate, it is assumed that a percentage of the crust is anorthositic. An additional assumption is the fractionation of $Al_{2}O_{3}$ that went into various minerals that crystallized. Some works have assumed that all of the $Al_{2}O_{3}$ went into forming plagioclase feldspar \citep[e.g.][]{Warren_1985}, while others have assumed that some of the $Al_{2}O_{3}$ also went into forming spinel \citep{ElkinsTanton_2011}. Due to the varying assumptions, the estimated LMO initial depths range from 100 to 1000~km \citep{Hodges_1974, Walker_1975, Solomon_1976, Solomon_1980, Kirk_1989, ElkinsTanton_2011, AndrewsHanna_2013, Lin_2017b}. Similarly, the solidification time also has a range of estimates. For thermal models, it ranges from 10 to nearly 300~Myrs \citep{Solomon_1977b, Minear_1980, Meyer_2010, ElkinsTanton_2011}, while for geochemical analyses, it ranges from about 100 to 254~Myrs \citep{Nyquist_1995, Rankenburg_2006, Boyet_2007, Nemchin_2009}. To be consistent with the LMO model, it is important that the solidification time of the LMO is comparable to the time span of primordial lunar crustal ages. Yet, this does not seem to be the case. Recent work suggested the LMO would have crystallized in 10~Myrs \citep{ElkinsTanton_2011}, which is much faster than what is suggested by the range of crust sample ages of $\sim$~200~Myrs \citep{Alibert_1994, Borg_1999}. Further work is required to refine these ages to be consistent with each other.

\subsection{Re-impacting Debris}
\label{sec:Introduction:Debris}

Recent work has shown that a substantial amount of debris (about $10^{23}$~kg or $\sim$~1.3 lunar masses) had sufficient speed to escape the Earth--Moon system after the Moon forming impact \citep{Kokubo_2000b, Kokubo_2000, Marcus_2009, Jackson_2012}. That quantity of escaping debris is for the Canonical model, which is a rather gentle impact with an impact velocity only just above the escape velocity. For many of the newer, modified versions of the Giant Impact Model, the giant impacts are more violent, thus they tend to produce more escaping debris. \citet{Leinhardt_2012} find that a typical giant impact releases around 3 to 5\% of the colliding mass as debris, as compared to 1.6\% for the Canonical model. For this work, we are making a conservative estimate by assuming the quantity of debris is that of the Canonical model.

While on heliocentric orbits, much of the debris would have subsequently re-impacted onto both the Earth and the Moon \citep{Daly_1946, Jackson_2012}. \citet{Jackson_2012} found that within a million years after the giant impact, debris would have accreted onto the Moon at an average rate of $\sim$~$9 \times 10^{13}$~kg/yr (with 50\% loss to collisional grinding). Debris would have re-impacted the Moon while the LMO was solidifying. Impacts could have significantly altered the cooling rate when the Moon had developed a conductive lid (i.e. at the point of plagioclase stability). Impacts that punctured holes into a conductive lid would have increased the thermal flux by exposing magma that used to be thermally insulated. A similar scenario is expected on Europa when impacts puncture holes into its ice shell to expose liquid water beneath \citep{Bauer_2011}. \citet{Hartmann_1980} proposed that early impacts should have pulverized the nascent floatation crust and they may have sped up the LMO solidification; however, they did not quantify the LMO solidification time. \citet{Minear_1980} and \citet{Davies_1982} both argued that impacts should have sped up LMO solidification; however, their models were highly simplified.

\subsection{Scope of this Work}
\label{sec:Introduction:Scope}

In this work, we include the sustained bombardment of debris generated after the giant impact with the thermal evolution of the LMO. We are primarily interested in how re-impacting debris affects the thermal evolution of the LMO. For our work, we use a model that can thermally evolve the LMO while producing and thermally evolving holes generated in the lunar crust by re-impacting debris. In Section \ref{sec:Methods:Debris} we discuss numerical calculations of debris evolution and in Section \ref{sec:Methods:Code} we discuss the details of our thermal evolution code. In Section \ref{sec:Results} we show our results. In Section \ref{sec:Discussion} we discuss consistency of our results with lunar crust sample ages, implications for the lunar surface and interior, and implications for the lunar orbital evolution.

% METHODS
\section{Methods}
\label{sec:Methods}

Since we were interested in the bulk, rather than spatially resolved, properties of the LMO thermal evolution, for this work we use a 1-D spherically symmetric thermal model. Following a similar procedure to \citet{ElkinsTanton_2011}, we use minute volume segments to iteratively solidify the modeled LMO and release the relevant energy through the modeled Moon's surface. Unlike previous work, here we consider the effect of re-impacting debris on the solidification of the LMO. In Section \ref{sec:Methods:Code} and in following subsections we discuss the thermal evolution code in detail; however, we begin with Section \ref{sec:Methods:Debris} by discussing the expected quantity and evolution of debris after the Moon forming impact.

\subsection{Re-impacting Debris Evolution}
\label{sec:Methods:Debris}

As noted above, the Canonical Moon-forming impact results in the release of around 1.3 lunar masses of material onto heliocentric orbits \citep[][]{Jackson_2012}. That mass is comparable to the mass that remains in Earth orbit as the proto-lunar disk \citep[e.g.][]{Canup_2004}. As it orbits the Sun this debris will encounter the terrestrial planets, especially Earth since it by definition must begin on Earth-crossing orbits, and will be re-accreted over time. \citet{Jackson_2012} conducted an extensive analysis of the dynamical evolution of the heliocentric Moon-forming debris using $N$-body simulations. We utilize the results of an improved $N$-body simulation that uses the same initial conditions and setup as \citet{Jackson_2012} and the same \textsc{Mercury} integrator \citep{Chambers_1999}, but with an increased number of debris particles ($10^{5}$) and a longer integration time of 100~Myr rather than 10~Myr.

As described by \citet{Jackson_2012} it is not feasible to resolve the orbit of the Moon in a long term dynamical simulation and as such the Earth and Moon are treated as a single body.  The debris accretion rate determined from the simulation is thus the accretion rate onto the Earth-Moon system as a whole.  To separate them we need to know the ratio between the accretion rates for Earth and the Moon.  \citet{Bandermann_1973} derive an analytic relation for the accretion ratio between the Earth and Moon, which they give as
\begin{equation}
\frac{A_{\rm E}}{A_{\rm M}} = \left(\frac{R_{\rm E}}{R_{\rm M}}\right)^2\frac{1+u^2}{\frac{7}{6}\frac{R_{\rm E}}{r} + 0.045 + u^2},
\label{eq:bandermann}
\end{equation}
where $A_{\rm E}$ and $A_{\rm M}$ are the accretion rates for Earth and the Moon, $R_{\rm E}$ and $R_{\rm M}$ are the respective radii, $r$ is the
Earth-Moon separation and $u$ is the ratio of the relative velocity to the escape velocity of Earth, $v_{\rm rel}/w_{\rm E}$.  Note that strictly this is a lower limit to the accretion ratio (or an upper limit to the lunar accretion rate) since it ignores the effect of shadowing by Earth, but this effect is small at all but the smallest Earth-Moon separations.  The impact velocity, $v_{\rm imp}$, of each impacting debris particle is provided by the $N$-body simulation and the relative velocity is then just $v_{\rm rel} = \sqrt{v_{\rm imp}^2-w_{\rm E}^2}$.

\begin{figure}[ht]
\includegraphics[width=0.996\textwidth]{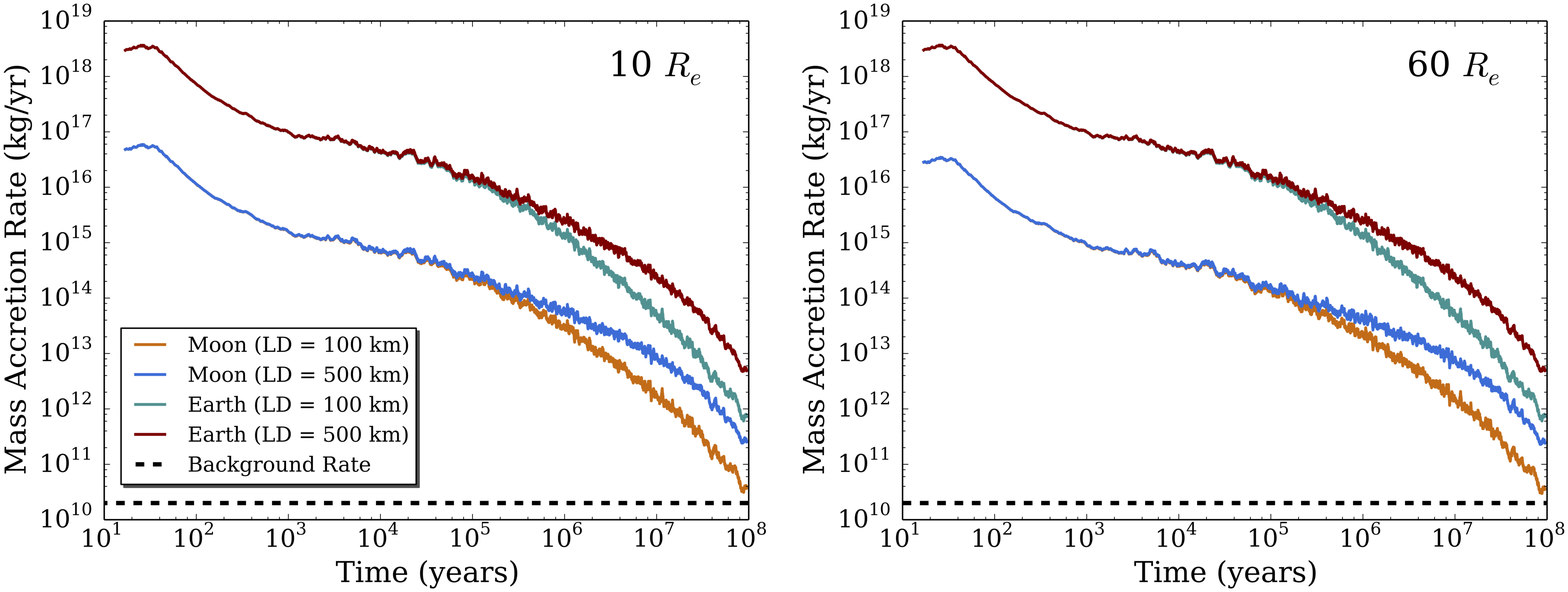}
\centering
\caption{Mass accretion rate over time for the Earth (dark cyan and red lines) and the Moon (blue and orange lines) for two populations of re-impacting debris based on the size of the largest debris (LD). On the left is the accretion rate when the Moon is at a distance of 10 Earth radii and on the right when it is at a distance of 60 Earth radii. An estimate for the accretion rate due to `background' asteroidal impacts during the proposed Late Heavy Bombardment \citep{Ryder_2002} is shown by a black dashed line for comparison.}
\label{fig:Debris}
\end{figure}

In addition to the relative velocity that we can determine from the $N$-body simulation the accretion ratio also depends on the Earth-Moon separation, which is an independent parameter.  Today the Earth-Moon separation is 60~$R_{\rm E}$, however the Moon is migrating outwards over time and would likely have formed near the Roche limit at around 3~$R_{\rm E}$. The timeline of the evolution of the lunar orbit is complicated however, especially at early times, and as the Moon crossed orbital resonances it likely went through high-eccentricity periods that further complicate the picture \citep[e.g.][]{Touma_1998}.  Furthermore, the thermal state of the Moon and the rate of tidal evolution are somewhat coupled, as studied by \citet{Tian_2017}, such that if we expect the thermal evolution of the Moon to change as a result of re-impacting debris, this would also change the tidal evolution.  Nonetheless, the Moon likely reached a separation of 10~$R_{\rm E}$ quite rapidly \citep[e.g.][]{Touma_1994,Touma_1998}, and beyond this the accretion ratio changes fairly slowly (see Figure \ref{fig:Debris}). As such we use a constant Earth-Moon separation of 10~$R_{\rm E}$ as being relatively representative of the early Moon.  As shown in Figure~\ref{fig:Debris} the difference between the accretion rates at 10~$R_{\rm E}$ and 60~$R_{\rm E}$ is very small and so the exact distance assumed is not very important.

While the $N$-body simulation in combination with Equation~\ref{eq:bandermann} provides us with the rate at which the massless $N$-body debris particles strike the Moon, we need to convert this into a mass accretion rate. Individual bodies in the disk of heliocentric debris will collide with one another and gradually break up into ever smaller fragments until the resulting dust is small enough (roughly 1~$\mu$m) that it can be removed from the Solar System by radiation pressure. To calculate an accurate mass accretion rate we must account for this evolution of the debris through self-collision between debris fragments.

To compute the collisional evolution we use the code developed by \citet{Jackson_2014}, which improves on that of \citet{Jackson_2012}, accurately accounting for the initial asymmetry in the debris disk and allowing the mass assigned to each $N$-body particle to evolve individually. The collisional evolution is dependent on the size-distribution of the debris fragments.  The shape of the size-distribution is poorly constrained and so we follow \citet{Jackson_2012} and \citet{Jackson_2014} and make the assumption that the size-distribution is a single power law \textcolor{black}{such that the number of bodies with diameters between $D$ and $D+dD$, $n(D)dD$, is proportional to $D^{-7/2}dD$.} This is the slope to which a self-similar collisional cascade will relax over time \citep[e.g.][]{Dohnanyi_1969, Tanaka_1996}.  Small fragments will have short collisional lifetimes and thus will rapidly evolve towards collisional equilibrium, where this slope will be a relatively accurate reflection of the reality \citep[e.g.][]{Wyatt_2011}. Larger, longer lived fragments will evolve more slowly and so this assumption is less certain, however we have no evidence to support a different distribution and this is the simplest assumption.

For a size distribution that is in collisional equilibrium the evolution of the mass in the cascade is governed by the size of the largest fragments, since these are the longest lived, meaning that as these largest bodies break up their mass is redistributed down the cascade on timescales short compared with their lifetimes. Mass is ultimately lost from the cascade once it reaches micron sizes at which point Solar radiation pressure is sufficient to blow the dust out of the Solar System. This is a very useful property as it means that in addition to the assumption of the shape of the size-distribution we need to make only one further assumption, which is the size of the largest objects in the debris. In Figure~\ref{fig:Debris} we show the evolution of the mass accretion rate for two different values of the size of the largest object in the debris, 500~km and 100~km. \textcolor{black}{Smaller objects undergo collisional evolution more rapidly and thus the distribution in which the largest object is 100~km in diameter loses mass faster than that in which it is 500~km, resulting in lower accretion rates at later times.  \citet{Jackson_2012} argued that objects much larger than around 500~km are implausible for the Moon-forming impact and chose this as their fiducial estimate of the size of the largest objects.  They noted however that the true size of the largest objects is highly uncertain and tested a very wide range of values.}  \textcolor{black}{As such} we choose to use 100~km as the size of the largest object in the debris for our study \textcolor{black}{as a somewhat conservative estimate, but note that the true value could be somewhat higher or lower.  As we can see from Figure~\ref{fig:Debris} this results in an uncertainty of roughly one order of magnitude in the accretion rate at late times}. It is important though to note that while we have had to make these assumptions about the size-distribution to determine the rate of collisional evolution of the debris this only feeds into our work through the mass accretion rate, it does not influence any of the aspects of our study.

\subsection{Thermal Evolution Code}
\label{sec:Methods:Code}

As stated earlier, there are many estimates for the initial LMO depth; however, no geochemical modeling estimate has called for an entirely molten early Moon. Additionally, \citet{Salmon_2012} have argued that a cold ``parent body'' (about 40\% of lunar mass), which would not have undergone intense bombardment, would have formed shortly after the giant impact. Thus, here we assume that the Moon formed with a nearly solidified interior and a molten exterior \citep[e.g.][]{Solomon_1986}. We use a nominal LMO depth of 1000~km similar to \citet{ElkinsTanton_2011} for our work and in \textcolor{black}{Appendix \ref{Apd:Sensitivity}} we show that the LMO solidification time is rather insensitive to the initial depth. We recognize that the Moon could have a liquid outer core \citep{Williams_2014, Matsuyama_2016}; however, provided it is not undergoing significant solidification (and attendant heat loss), that should be inconsequential to the LMO's overall thermal evolution. We also ignore the lunar core formation \textcolor{black}{(differentiation)} process since \citet{Solomon_1980} showed that it only raised the average temperature of the Moon by 10~K.

Unlike, for example, \citet{Minear_1978} and \citet{ElkinsTanton_2011}, we did not explicitly model the geochemical crystallization of the LMO. We are interested in how re-impacts affected the bulk thermal properties of the LMO (e.g. its overall solidification time), rather than the geochemical internal structure of the Moon. This choice simplified the code and made it faster, which allowed us to explore a wider parameter space to better understand the effect of re-impacts. This is further justified by \citet{Minear_1978}, who found that the LMO solidification time is mainly dependent on the mode of heat transportation, crust thermal conductivity, and final crustal thickness. A limitation of not modeling the geochemical crystallization process is that there is not a natural prescription for what fraction of the solidified material should sink or float. To approximate how crystallizing material would partition according to density, initially, for each iterative step, all material that solidified is assumed to be denser than the LMO and thus is added to the top of the solid interior. To model floatation crust formation, when the LMO depth decreases to 100~km, instead of all crystallizing material sinking to the interior, a fixed fraction (viz. 45\%) is directed to the surface to form crust. This is similar to \citet{Tian_2017} who used a 40\% fraction in the final 110~km of the LMO with the crust beginning with a thickness of 5~km. We choose this partitioning fraction to closely replicate the geochemical evolution in \citet{ElkinsTanton_2011} and the current crustal thickness of the Moon \citep{Wieczorek_2013}.

\textcolor{black}{The initial LMO volume is divided up into a user defined number of equal segments.} Our Python code then iterates over these minute LMO volume segments. At each iteration, the total energy released through the modeled surface is the sum of energy released due to secular cooling and partial solidification of the LMO. The total energy is allowed to be released both via direct thermal radiation \textcolor{black}{(from a free magma surface) or} conduction \textcolor{black}{(through an insulating layer of crust)} depending on the surface conditions. Unlike \citet{ElkinsTanton_2011}, we do not assume that an early atmosphere would be capable of maintaining a free, liquid surface to the LMO. Instead, we allow quench crust to form if the conditions are suitable (see Section \ref{sec:Methods:Code:Quench}). Quench crust is a rapidly solidified layer of crust that is approximately the same composition as the liquid magma. Therefore, in this work, the LMO cooling is initially controlled by thermal conduction through the quench crust. 

The temperature in the LMO is estimated by calculating the temperature at the solid-liquid boundary at the base of the LMO and \textcolor{black}{then} using the adiabat slope to calculate the temperature \textcolor{black}{throughout the rest of the LMO}. By definition, the temperature at the solid-liquid boundary will be the solidus temperature of the LMO at the relevant depth/pressure. We use the same solidus temperature equation (Equation \ref{eq:SolidusTemp}) as in \citet{ElkinsTanton_2011}. The solidus temperature (in Kelvin) is given by
\begin{equation} 
\label{eq:SolidusTemp}
T_s(r) = \Big(-1.3714 \times 10^{-4}\Big) r^{2} - 0.1724r + 2134.15 - \frac{4.4}{0.2L + 0.01},
\end{equation}
where $r$ is the radial position from the Moon's center in km and $L$ is the remaining liquid fraction of the LMO ranging from 1 to 0. The surface temperature is calculated self-consistently by equating the conductive flux through the crust to the radiative flux on the surface as given by
\begin{equation} 
\label{eq:SurfTemp}
\kappa_{c} \rho_{c} c_{c} \bigg (\frac{T_{bc} - T_{tc}}{d_c} \bigg) = \epsilon \sigma \Big(T_{tc}^{4} - T_{e}^{4} \Big),
\end{equation}
where $\kappa_{c}$ is the thermal diffusivity of the crust, $\rho_{c}$ is the density of the crust, $c_{c}$ is the specific heat capacity of the crust, $T_{bc}$ and $T_{tc}$ are the temperatures at the bottom and at the top of the crust respectively, $d_{c}$ is the crustal thickness, $\epsilon$ is the emissivity, $\sigma$ is the Stefan--Boltzmann constant, $T_e$ is the equilibrium temperature of the surface in the absence of internal heat sources, \textcolor{black}{and we assume that the material properties of the crust are indepedent of depth and temperature}. At each iterative step, we solve for the surface temperature (i.e. $T_{tc}$) that equates the conductive and radiative fluxes.

Whether quench or floatation crust is present on the surface, the simple conductive energy release is complicated by re-impacting debris, which may puncture holes into the crust. At each iteration, we consider the area of holes that are punctured by impacts (see Section \ref{sec:Methods:Code:Impacts}). We calculate the equilibrium quench crust that would form in a particular hole and we account for the increased conductive flux (due to the thin layer of quench) in the energy release calculations.  \textcolor{black}{The surface temperature (as given by Equation~\ref{eq:SurfTemp}) and the heat flux from the surface are thus not calculated on a global basis, but on a local basis to account for the varying thickness of the crust.  The total rate of heat loss is then determined by multiplying the fluxes by the relevant areas.}

Since we use increments of constant volume rather than constant time, it is necessary to calculate the time taken for each volume increment to solidify, which is simply the energy that must be released in that step divided by the net heat flux at the lunar surface.  Note that the quantity of material accreted, and thus the area of holes produced during the solidification of a volume increment is dependent on the time taken, but that the area of holes produced will also influence the time taken.  As such within the calculation for each volume increment an iteration is required to ensure consistency.  While this adds to the computational cost of each volume increment calculation it converges quickly and the very large variation in solidification rates over a complete run makes this preferable to using increments of constant time.

The iteration is terminated when 1\% by volume of the initial LMO remains. Previous work, such as \citet{ElkinsTanton_2011}, typically stop their calculations at that point since the remaining liquid consists of incompatible elements and is proposed to be the ur-KREEP layer (i.e. the hypothesized source region of KREEP elements on the lunar surface) \citep{Warren_1979}. We list the nominal values for the relevant parameters used for these calculations in Table \ref{tableConstants}.

During the thermal evolution of the LMO, it is likely that it had additional heat sources due to some or all of the following: secular cooling of the core \citep[e.g.][]{Konrad_1997}, radiogenic heating \citep[e.g.][]{Meyer_2010, ElkinsTanton_2011}, and tidal heating \citep{Meyer_2010, Chen_2016}. In this work, we do not explicitly consider individual heat sources but we allow for additional energy to be added to the LMO during its thermal evolution (see Section \ref{sec:Discussion:CrustAge}).

%electrical induction heating \citep{Sonett_1975}

\begin{table}
\caption{Nominal Parameter Values}
\centering
\begin{tabular}{c c c p{1.5in} p{1.6in}}
Symbol & Value & Units & Description & Reference \\
\hline
\hline
$S_a$ & $1.5 \times 10^{-4}$ & K/m & Adiabat Slope & \citet{Zhang_2013} \\
$d_m$ & 1000 & km & LMO Initial Depth & \citet{ElkinsTanton_2011} \\
$d_f$ & 100 & km & Floatation Crust Formation Depth & \citet{ElkinsTanton_2011} \\
$f_p$ & 0.45 & -- & Floatation Crust Partition Fraction & Set to match lunar crustal thickness\\
$z_m$ & 1\% & -- & Residual LMO & \citet{Warren_1979} \\
$H_f$ & $4.187 \times 10^{5}$ & J/kg & LMO Heat of Fusion & \citet{ElkinsTanton_2007, Piskorz_2014} \\
$\alpha_m$ & $3 \times 10^{-5}$ & 1/K & LMO Thermal Expansion Coefficient & \citet{ElkinsTanton_2007} \\
$\eta_m$ & $10^{3}$ & Pa $\cdot$ s & LMO Dynamic Viscosity & \citet{Bottinga_1972} \\
$\rho_m$ & 3.0 & g/cm\textsuperscript{3} & LMO Density & \citet{Meyer_2010} \\
$\rho_c$ & 2.7 & g/cm\textsuperscript{3} & Crust Density & \citet{Gast_1972b} \\
$\rho_q$ & 2.7 & g/cm\textsuperscript{3} & Quench Density & Set equal to crust value for quench floatation \\
$c_m$, $c_c$, $c_q$ & 1256.1 & J/kg $\cdot$ K & LMO, Crust \& Quench Specific Heat Capacity & \citet{ElkinsTanton_2007, Eppelbaum_2014} \\
$\kappa_m$, $\kappa_c$, $\kappa_q$ & $10^{-6}$ & m\textsuperscript{2}/s & LMO, Crust \& Quench Thermal Diffusivity & \citet{ElkinsTanton_2007, Eppelbaum_2014} \\
$d_q$ & 10 & m & Maximum Quench Thickness & \citet{Rathbun_2002, Matson_2006} \\
$T_{melt}$ & 1000 & K & Quench Melting Temperature & \citet{Eppelbaum_2014} \\
$T_e$ & 250 & K & Equilibrium Radiative Temperature & Approximate lunar equilibrium temperature without an atmosphere \\
$\epsilon$ & 1.0 & -- & Emissivity & Idealized perfect emitter \\
$a_g$ & 1.6 & m/s\textsuperscript{2} & Acceleration Due to Gravity & Approximate surface value \\
\hline
\label{tableConstants}
\end{tabular}
\end{table}

\subsubsection{Quench Crust}
\label{sec:Methods:Code:Quench}

Quench crust at the early stage of the LMO's thermal evolution has been considered inconsistently, with some works having included it \citep[e.g.][]{Minear_1980}, while others having disregarded it \citep[e.g.][]{ElkinsTanton_2011}. Besides the Moon quench crust has also been considered for Mercury \citep{Riner_2009}, which may have also had a floatation crust form from a magma ocean \citep{VanderKaaden_2015}.

The choice of whether or not to include quench crust will influence the LMO solidification time.  Since a quench crust adds a conductive layer to the surface of the LMO it will reduce the heat flow out of the LMO and so increase the solidification time.  In the absence of impacts the question of quench crust affects only the early phase of LMO cooling before the formation of floatation crust begins.  While the length of this early phase can be affected considerably by the presence or absence of quench crust, it is always a small fraction of the total solidification time, as discussed in Section~\ref{sec:Results:Time} and so the influence on the total solidification time is negligible. When impacts are included the question of quench crust comes into play in the floatation crust phase as well, since our impacts punch through to the LMO and a hole that is covered by a layer of quench crust will \textcolor{black}{have} a reduced influence on the cooling rate compared to a hole in which magma is directly exposed. As such considering quench crust gives us a conservative estimate of the influence of hole production.

We argue that quench crust could have been present for two reasons. First, we compare the convective flux from the LMO to the radiative flux from the surface. If the LMO radiated directly to space with a surface temperature $>$~1000~K and an equilibrium temperature of 250~K, convection would not be able to deliver heat to the top of the LMO fast enough to balance the rate of heat loss by thermal radiation. As such, quench crust would form on the surface. This can be shown using the Nusselt number (\textit{Nu}), which is the ratio of convective and conductive heat fluxes and is given by 
\begin{equation} 
\label{eq:Nusselt}
Nu = a \cdot Ra^{\beta},
\end{equation}
where $a$ and $\beta$ are constants. We use \textit{a} = 0.124 and $\beta$ = 0.309 from experimental work by \citet{Niemela_2000} (see Appendix \ref{Apd:Nusselt} for additional details). \textit{Ra} is the Rayleigh number, which is given by
\begin{equation} 
\label{eq:Rayleigh}
Ra = \frac{a_g \cdot \rho_{m} \cdot \alpha_{m} \cdot \Delta T \cdot d_m^{3}}{\eta_{m} \cdot \kappa_{m}},
\end{equation}
where $a_g$ is acceleration due to gravity, $\rho_{m}$ is the density of the LMO, $\alpha_{m}$ is the thermal expansion coefficient of the LMO, $\Delta T$ is the temperature difference, $d_m$ is the depth of the LMO, $\eta_{m}$ is the dynamic viscosity of the LMO, and $\kappa_{m}$ is the thermal diffusivity of the LMO. Using an initial $\Delta T$ of 150~K (adiabatic temperature change over 1000~km from the base of the LMO to the base of the quench crust) along with nominal values from Table \ref{tableConstants}, \textit{Ra} is approximately $2 \times 10^{22}$ and in turn, \textit{Nu} is approximately $10^{6}$ \textcolor{black}{(note that as discussed in Appendix~\ref{Apd:Nusselt} this is subject to a fair degree of uncertainty)}. The conductive heat flux of the LMO is given by
\begin{equation} 
\label{eq:CondFluxMO}
\textcolor{black}{F_{cond}^{m}} = \kappa_{m} \rho_{m} c_{m} \frac{T_{mb} - T_{mt}}{d_{m}},
\end{equation}
where $\kappa_{m}$ is the thermal diffusivity of the LMO, $\rho_{m}$ is the density of the LMO, $c_{m}$ is the specific heat capacity of the LMO, $T_{mb}$ and $T_{mt}$ are the temperatures at the bottom and at the top of the LMO respectively, and $d_{m}$ is the thickness of the LMO. Using our nominal values along with initial values for $T_{mb}$ equal to 1912~K (the solidus temperature at the initial solid-liquid boundary) and $T_{mt}$ equal to 1400~K (the solidus temperature at the top of the LMO initially, which we thus expect to be the temperature at the base of the quench layer), the conductive flux is equal to $\sim$~$2 \times 10^{-3}$~W/m\textsuperscript{2}. By using \textit{Nu}, we calculate the convective flux to equal $\sim$~$2 \times 10^{3}$~W/m\textsuperscript{2}. This is about two orders of magnitude smaller than the radiative flux of a surface with a temperature of 1400~K and an equilibrium temperature of 250~K (i.e. $\sim$~$10^{5}$~W/m\textsuperscript{2}). Therefore, a substantial atmosphere would be required to decrease the radiative flux from the surface and thus to prevent quench crust formation.

We will now consider the plausibility of a thick early lunar atmosphere. It is possible that an early lunar atmosphere was generated by vapor outgassed by the LMO and/or water released by impacts. That was likely the case for the Earth, where an early steam atmosphere may have kept the surface from rapidly solidifying (i.e. forming quench crust) \citep{Abe_1986}. The Moon, however, is depleted in volatiles relative to Earth \citep{Taylor_2014}, and being less massive has a significantly larger surface area to mass ratio such that any atmosphere will be spread more thinly. Furthermore, the Moon formed at approximately the Roche limit \citep{Canup_2004}. Thus, any initial atmosphere would have been highly susceptible to Roche lobe overflow \citep[e.g.][]{Repetto_2014}, especially considering the large scale height a hot early lunar atmosphere would have had. There are also other depletion mechanisms to consider including hydrodynamic escape \citep[e.g.][]{Pepin_1991}, impact removal \citep[e.g.][]{Melosh_1989} and charged particle interactions \citep[e.g.][]{Luhmann_1992}. Thus, it seems unlikely that the Moon was able to retain a substantial atmosphere for at least 1,000 years (the approximate time required to start forming floatation crust according to \citet{ElkinsTanton_2011}). As such, we consider quench crust to have been present atop the LMO.

We use the work of \citet{Matson_2006} for Loki Patera on Io as a model for quench crust formation and evolution. Their model considered Loki Patera to be a silicate `magma sea' that is large enough to have negligible shore influence and deep enough to ignore floor effects. That model should be readily extensible to the LMO in which there were no shores and which was deep. \citet{Matson_2006} also argued that though material that solidifies out of magma is generally denser than the magma itself and thus should sink \citep[e.g.][]{Walker_1980, Walker_1985, Spera_1992}, solidified material should be buoyant due to trapped volatiles until it grows to a certain maximum thickness. As solidified material gets thicker, additional material would be solidifying at greater depths meaning that the bubbles would be smaller and provide less buoyancy. As such \citet{Matson_2006} argued at greater than approximately 7~m thickness (for magma on Io), the solidified layer should sink. Similarly, quench crust thicknesses have been estimated to be about 6~m for Hawaiian lava lakes \citep{Rathbun_2002}. This suggests that maximum quench crust thickness in the order of 10~m may be ubiquitous to silicate magmas on any planet. Hence, for our work, we allow quench crust to \textcolor{black}{grow} up to a nominal maximum thickness of 10~m. If sufficient volatiles are not incorporated into quench crust, quench crust should sink due to its higher density compared to the LMO. That would make the previous discussions regarding the flux balance and the early lunar atmosphere immaterial. While there are uncertainties regarding quench crust, we will show below that it has nearly no effect on the overall LMO solidification time.

Equilibrium quench crust thickness is calculated by equating the convective heat flux out of the LMO, the conductive heat flux through the quench crust, and the radiative flux from the top of the quench crust. Once the convective heat flux of the LMO has been calculated, we then calculate the temperature at the top of the quench crust by setting the convective heat flux from the LMO equal to the radiative heat flux from the quench crust surface using
\begin{equation} 
\label{eq:TempTopQuench}
T_{tq} = \Big(\frac{\textcolor{black}{F_{conv}^{m}}}{\epsilon \sigma} + T_{e}^{4} \Big)^{1/4},
\end{equation}
where  $T_{tq}$ is the temperature on top of the quench crust, \textcolor{black}{$F_{conv}^{m}$} is the convective heat flux of the LMO, and $T_{e}$ is the equilibrium temperature of the atmosphere. $T_{tq}$ is usually close to the radiative equilibrium temperature of 250~K. With the temperatures at the bottom and at the top of the quench crust defined, the thickness of the quench crust can then be \textcolor{black}{calculated} by setting the conductive flux through the crust equal to the convective flux of the LMO as given by
\begin{equation} 
\label{eq:QuenchThickness}
d_q = \kappa_{q} \rho_{q} c_{q} \frac{T_{melt} - T_{tq}}{\textcolor{black}{F_{conv}^{m}}},
\end{equation}
where $d_q$ is the thickness of the quench crust, $\kappa_{q}$ is the thermal diffusivity of quench crust, $\rho_{q}$ is the density of quench crust, $c_{q}$ is the specific heat capacity of quench crust, and $T_{melt}$ is the melting temperature of quench crust (set to 1000~K). As stated earlier, at the start of LMO solidification we calculate the convective flux to equal $\sim$~$2 \times 10^{3}$~W/m\textsuperscript{2}. Additionally, at the point when floatation crust starts to form, the convective flux is equal to $\sim$~$1 \times 10^{2}$~W/m\textsuperscript{2}.

Quench crust growth rate may be quantified using the Stefan problem. However, since our timesteps are larger than the time required to form quench crust, we do not explicitly calculate quench growth at each iteration. Additionally, as stated above an early lunar atmosphere may have affected quench crust stability. Though we do not explicitly model an early lunar atmosphere, our formulation is sufficiently general to indirectly mimic an atmosphere (using emissivity and equilibrium radiative temperature).

\subsubsection{Incorporating Re-impacts}
\label{sec:Methods:Code:Impacts}

At each iterative step, our code looks up the mass of debris that impacted onto the Moon during that timestep using the results mentioned in Section \ref{sec:Methods:Debris}. As stated there, the mass of debris can be accurately quantified; however, the size distribution of the impactors is difficult to estimate due to the lack of constraints. In addition, even if a certain size distribution is assumed, it is unclear how hole diameter is related to the size and the velocity of an impactor. The diameter of a hole is also likely dependent on the crustal thickness and the strength of the crust, with both changing over time. As such, rather than attempting to model the production of individual holes at each timestep, we utilize a conversion factor, $k$, which is defined so that

\begin{equation} 
\label{eq:k}
A_{holes}(t_{step}) = \frac{M_{imp}(t_{step})}{k},
\end{equation}

where $A_{holes}(t_{step})$ is area of holes produced on the surface during a certain timestep and $M_{imp}(t_{step})$ is impacting mass during that same timestep. Thus, $k$ has units of kg/m\textsuperscript{2}. This method allows us to characterize bulk properties of the thermal evolution of the LMO without making assumptions about the impactor size or velocity distributions or the process of hole production.

Very small values of $k$ are unrealistic. For instance if $k = 10^{3}$~kg/m\textsuperscript{2}, that would mean that $\sim$~$4 \cdot 10^{16}$~kg of accreted mass (equivalent to a $\sim$~30~km object with a density of 3 g/cm\textsuperscript{3}) would produce an area of holes that is equivalent to the surface area of the Moon. For the lowest $k$ value used in this work (i.e.  $k = 10^{5}$~kg/m\textsuperscript{2}), $\sim$~$4 \cdot 10^{16}$~kg of accreted mass would produce an area of holes equal to about 1\% of the lunar surface area or a single hole with a radius of about 360~km. For this example, if the crustal thickness was 4~km, it means that impacting material is able to displace crustal mass about two orders of magnitude more than its own mass. On the other hand, larger values of $k$ means that, though the Moon is accreting mass by impacts, the conditions are such that few to no holes are produced, with the no holes case represented by the limit where $k~\rightarrow~\infty$. This could be due to small impactors, low-velocity impactors, a thick crust, and/or a high strength crust. For our largest $k$ value extremum we use $k = 10^{9}$~kg/m\textsuperscript{2}. In this case, only a small area of holes, equivalent to a single hole with a radius $\sim$~4~km, is produced by our hypothetical 30 km impactor, which is again an unlikely scenario. To estimate what might be typical values, we notice that the final lunar crust is around 45 km thick, and so thicknesses of around 10 km will be typical during LMO solidification. It seems reasonable to expect that a 10 km impactor could produce a hole of at least 10 km diameter in 10 km thick crust, which would correspond to $k\sim10^7$~kg/m\textsuperscript{2}. As such, our lower and upper bounds cover the range between a very intense bombardment that produces large hole areas and a very feeble bombardment that produces small hole areas, while we expect that values of $k$ around $10^7$~kg/m\textsuperscript{2} may be typical. \textcolor{black}{As a caveat we note that we should expect the value of $k$ to vary with time.  We would expect the increase in crust thickness over time to result in an increase in $k$, however the typical impact velocities also increase over time, which we would expect to decrease $k$.  We expect that the increasing crustal thickness will win out, but the exact behaviour we should expect of $k$ over time is not clear.}

\subsubsection{Distribution and Redistribution of Crustal Material}
\label{sec:Methods:Code:Crust}
Though we do not explicitly model individual impacts, there are certain physical effects that need to be implemented in the code to make the calculations realistic. One such effect is to allow holes to be closed naturally by newly formed floatation crust. This is implemented by dividing newly formed floatation crust material between existing crust and holes.  \textcolor{black}{For the majority of this work the division is simply} according to the surface area covered by each, \textcolor{black}{such that we can envisage plagioclase rising globally uniformly and joining the base of the existing crust, or filling in the holes}.  \textcolor{black}{In Section~\ref{sec:Results:Concentrating} however, we investigate the effect of concentrating the newly formed crust into the holes.}

The other effects that need to be considered are allowing impacts to occur in both non-impacted and previously impacted areas of the Moon and the conservation of crustal material. To account for allowing impacts in both non-impacted areas and previously impacted areas, when a new hole area is generated it is divided between areas that do and do not contain holes according to the surface area of the Moon covered by each so that a uniform probability of impact at any point on the lunar surface is maintained.  To ensure conservation of crustal material, when new hole area is generated the volume of crust removed from the new holes is spread \textcolor{black}{uniformly} across the remainder of the lunar surface.  \textcolor{black}{While for any individual impact basin it would be expected that the ejecta would be concentrated nearby, as we described previously our hole areas do not correspond to individual geographically localized holes, but rather to the global hole area generated during one timestep.}  \textcolor{black}{To ensure that we capture the infilling of the holes accurately rather than the surface simply being divided between `hole' and `not-hole' we track each newly generated area of holes individually after it is created.}

% RESULTS
\section{Results}
\label{sec:Results}

\subsection{Surface Area with Holes}
\label{sec:Results:SurfaceArea}

\begin{figure}[ht]
\includegraphics[width=0.80\textwidth]{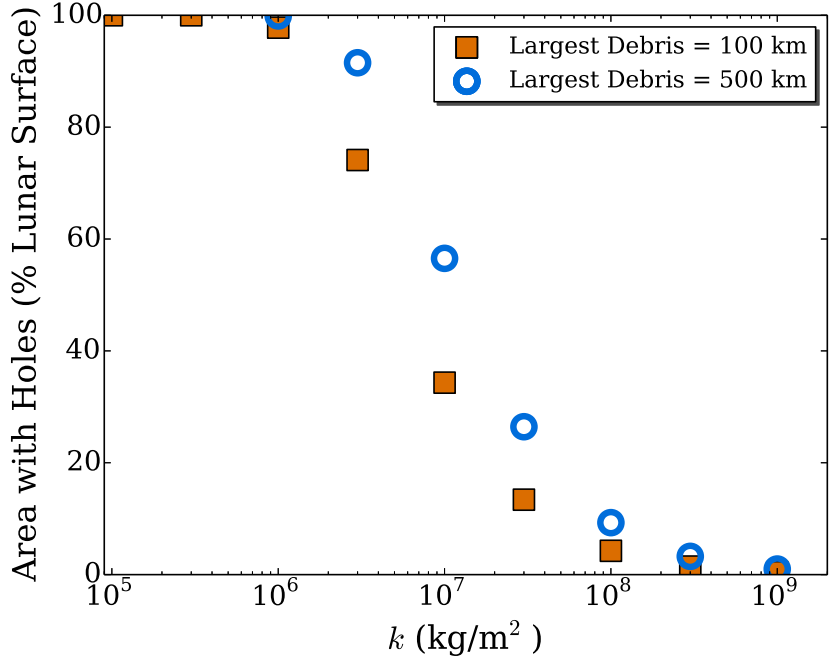}
\centering
\caption{Surface area of the Moon that has holes at the end of the iteration as a function of $k$. Cases where the largest debris is 100~km in size is shown with closed orange squares, while cases where the largest debris is 500~km in size is shown with open blue circles.}
\label{fig:AreaWithHoles}
\end{figure}

In Figure \ref{fig:AreaWithHoles} we show the effect of $k$ on the percentage of the Moon's surface that has been impacted at the end of the iteration. For large values of $k$ ($\sim$~$10^{9}$~kg/m\textsuperscript{2}), the percentage of the Moon with holes is close to zero and when $k$ is small ($\leq~10^{6}$~kg/m\textsuperscript{2}), nearly all of the Moon's surface had holes after the LMO solidified. Since both newly formed crustal material and crustal material removed from newly formed holes are distributed equally across the lunar surface, the crust thickness of holes will not catch up to the crustal thickness of non-impacted areas. Therefore, holes should theoretically be identifiable at the end of the LMO solidification since their crustal thicknesses will be less than non-impacted areas. In Figure \ref{fig:AreaWithHoles} we show results for two populations of debris, one with the largest object being 100~km in size (our nominal case) and one with the largest object being 500~km in size. For the 500~km case, the bombardment intensity does not decrease as rapidly as for the 100~km case (as shown in Figure~\ref{fig:Debris}), and as such, for a given $k$ value, the 500~km case produces more holes. The runs for the various values of $k$ had different numbers of volume segments to ensure convergence, as described in \textcolor{black}{Appendix ~\ref{Apd:Convergence}.}

\begin{figure}[ht]
\includegraphics[width=0.80\textwidth]{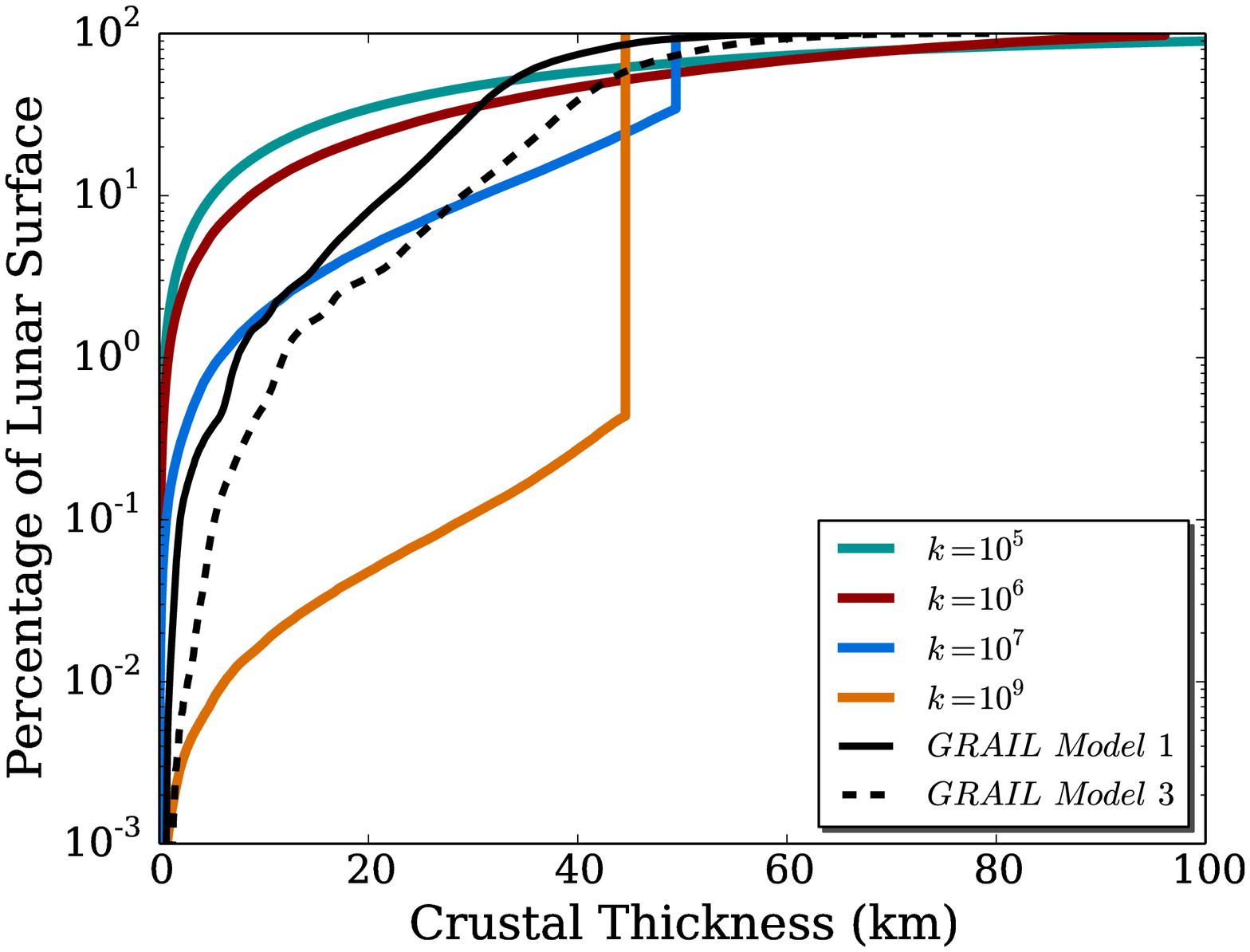}
\centering
\caption{Cumulative surface area of the Moon covered by crustal thickness of equal to or less than a particular thickness for different $k$ values (colored solid lines). Also shown are two crustal thickness models using GRAIL data by \citet{Wieczorek_2013} (solid and dashed black lines).}
\label{fig:CrustHisto}
\end{figure}

In Figure~\ref{fig:CrustHisto} we show the cumulative lunar surface area as a function of crustal thickness for different $k$ values at the end of the iteration. When $k$ is large, nearly all of the lunar surface consist of crust that is approximately the present mean lunar crustal thickness ($\sim$~45~km), as expected since in the absence of disturbances from hole production crust formation will proceed uniformly. For lower $k$ values, there is more of a distribution of crustal thicknesses. As a comparison we also show on Figure~\ref{fig:CrustHisto} Models 1 and 3 from \citet{Wieczorek_2013}, who used gravity data from the Gravity Recovery and Interior Laboratory (GRAIL) mission to produce improved lunar crustal thickness models. Both of these models assume a constant porosity of 12~percent throughout the lunar crust, but differ in the crustal thickness constraints used for the Apollo 12 and Apollo 14 landing sites (30~km for Model 1 and 38~km for Model 3), yielding mean crustal thicknesses of 34~km for Model 1 and 43~km for Model 3. While not a perfect match the \citet{Wieczorek_2013} models lie closest to the curve for $k=10^7$~kg/m\textsuperscript{2}. An exact match between the \citet{Wieczorek_2013} models and our distributions is not expected since our crustal thicknesses are those at the end of crust formation, and will be modified by billions of years of subsequent impact bombardment. Nonetheless the similarities between our distributions and the \citet{Wieczorek_2013} models suggest that some of the variation in crustal thicknesses may have already been in place very early in lunar history.  It is also reassuring that the greatest similarity is evident for $k\sim10^7$~kg/m\textsuperscript{2} since we noted in Section~\ref{sec:Methods:Code:Impacts} that this may be a typical value for $k$.

\subsection{Lunar Magma Ocean Solidification Time}
\label{sec:Results:Time}

\begin{figure}
\includegraphics[width=0.75\textwidth]{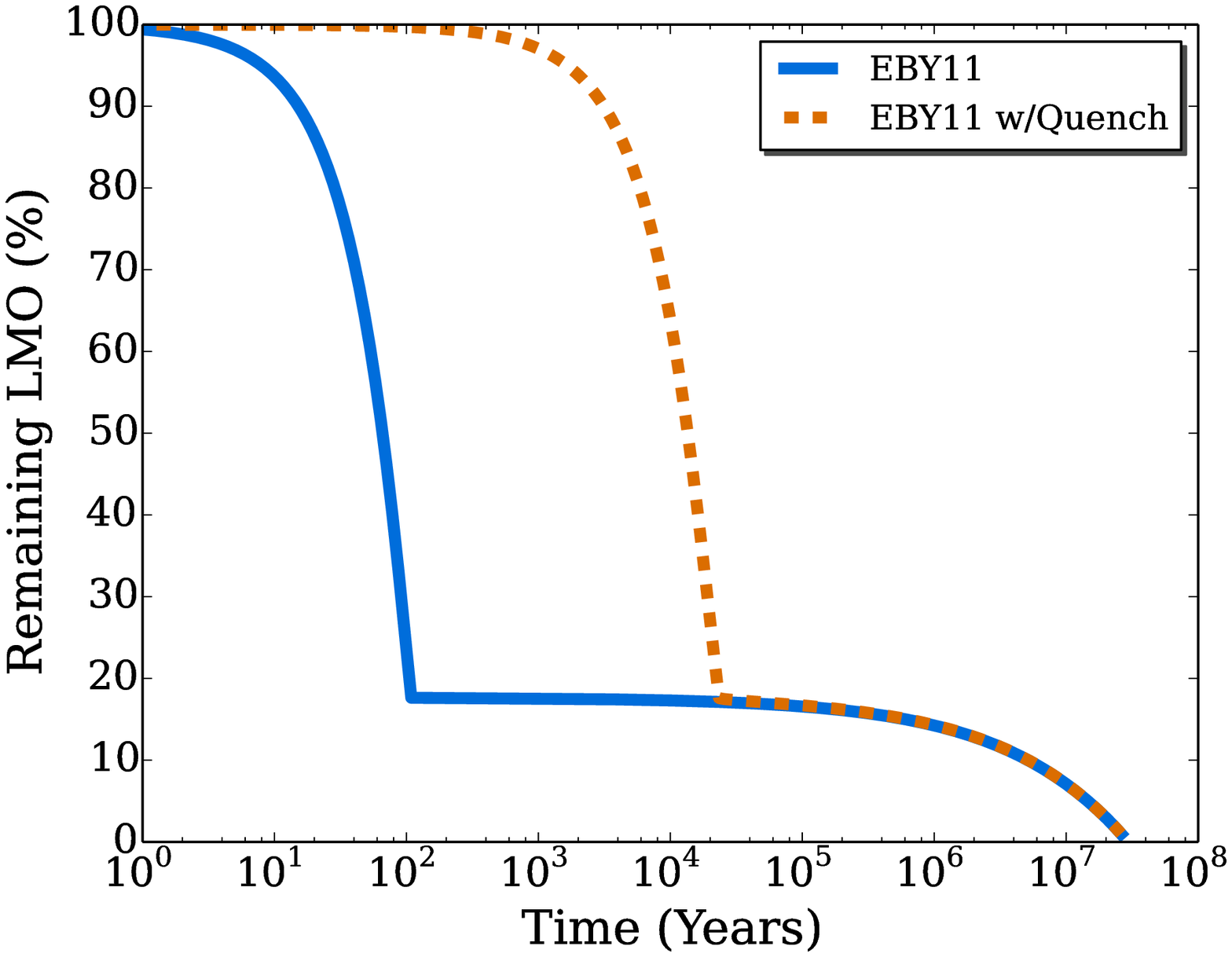}
\centering
\caption{Fraction of magma ocean remaining over time. The EBY11 cooling model is shown by the solid blue curve and the EBY11 model with quench crust is shown by the dashed orange curve. In the EBY11 model, the initial rapid cooling is due to the thermally radiative global surface, while the slower cooling from $\sim$~100~years onward is due to the presence of the thermally conductive global lid. The case with quench crust is similar to the EBY11 model, except the rapid cooling is delayed by $\sim$~$10^{4}$ years. Although nominally we used 250~K as the equilibrium radiative temperature, here we used a value of 25~K to approximate the temperature used in the EBY11 model.}
\label{fig:RemainingLiq}
\end{figure}

In Figure~\ref{fig:RemainingLiq} we show the fraction of the magma ocean remaining as a function of time for conditions matching those used by \citet{ElkinsTanton_2011}, i.e. a 25~K equilibrium radiative temperature, which we refer to as the EBY11 cooling model. We find a similar, but slightly longer solidification time of 26~Myr for these conditions, compared with the $\sim$~10~Myr found by \citet{ElkinsTanton_2011}. This difference is likely for of two reasons. Firstly, we did not explicitly model the fractional crystallization process like they do in their work. Secondly, here we assumed that the surface was conductive as soon as floatation crust started to form, rather than after 5~km of floatation crust had formed in their work. Nevertheless, the difference in solidification time is only a factor of $\sim$~2-3, thus it reassures us that we are capturing the broad characteristics of the cooling process and that we can realistically explore the effect of re-impacts on the cooling process. \citet{ElkinsTanton_2011} does not consider quench crust, rather keeping the liquid surface of the magma ocean exposed to space until plagioclase formation begins. For the solid blue curve in Figure~\ref{fig:RemainingLiq} we turn off the quench crust to replicate the fast early phase of \citet{ElkinsTanton_2011} in which the first 80\% of the LMO solidifies, which here we find takes around 100 years. Though quench crust does not significantly alter the overall solidification time of the LMO, it prolongs the early, rapid cooling phase by $\sim$~$10^{4}$~years, as shown by the dashed orange curve in Figure~\ref{fig:RemainingLiq}.

\begin{figure}
\includegraphics[width=0.75\textwidth]{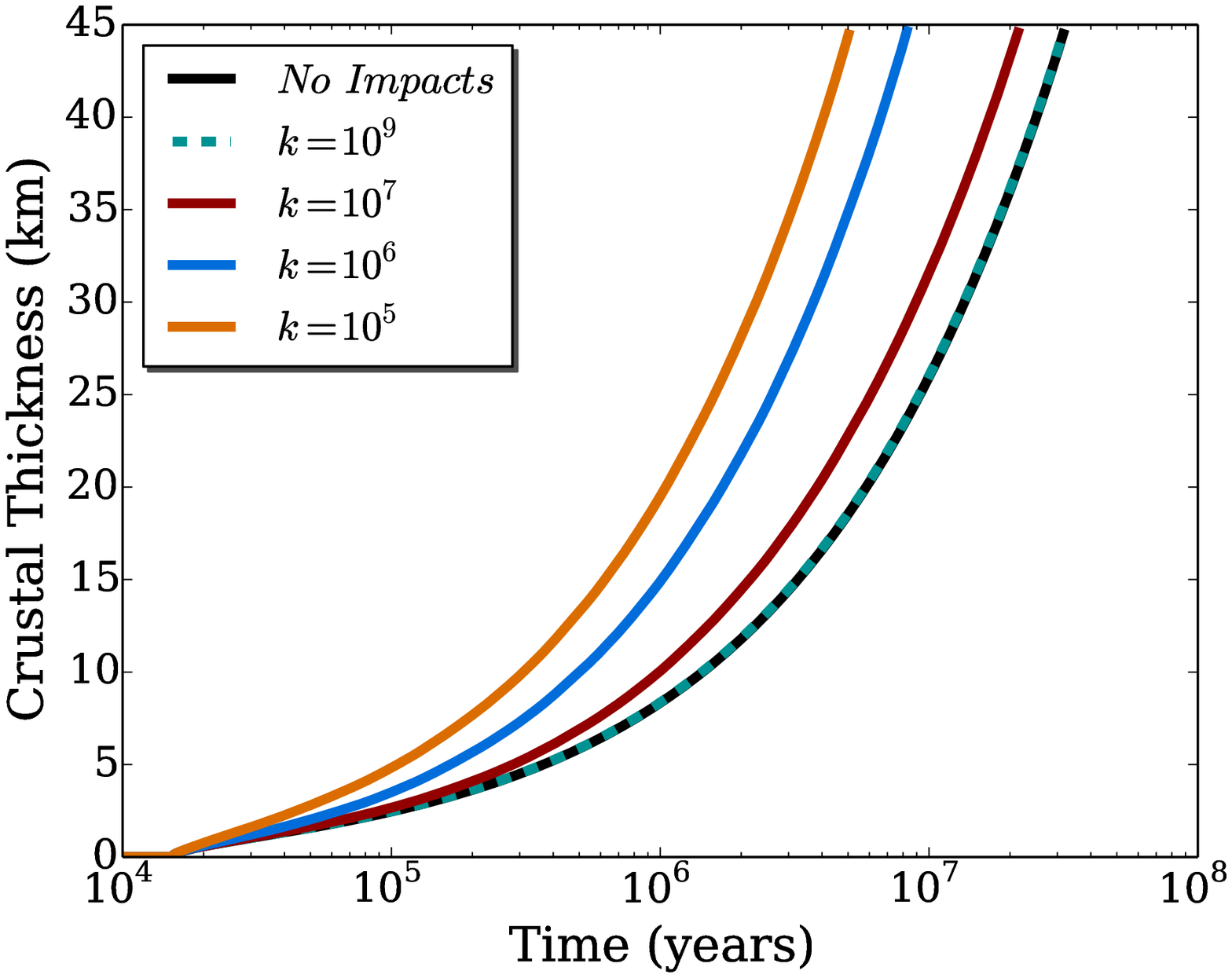}
\centering
\caption{Crustal thickness over time for different $k$ values (colored curves) compared to the no impacts case (black curve). The curve for the $k$~=~$10^{9}$~kg/m\textsuperscript{2} case is dashed so that the no impacts case is visible. All parameters aside from $k$ are set at their nominal values as given in Table~\ref{tableConstants}.}
\label{fig:CrustalThickness}
\end{figure}

In Figure \ref{fig:CrustalThickness} we show lunar crustal thickness over time as a function of $k$. As $k$ is reduced, the time taken for the completion of crust formation (and solidification of the LMO) decreases substantially, from $\sim$ 32~Myr at $k\geq10^9$~kg/m\textsuperscript{2} to only $\sim$ 5~Myr at $k=10^5$~kg/m\textsuperscript{2}. For $k\gtrsim10^9$~kg/m\textsuperscript{2}, the crust evolves as essentially identically to there being no holes generated by re-impacting debris. Nominally the number of volume segments used was $10^{5}$; however, for $k$~=~$10^{6}$~kg/m\textsuperscript{2} and $k$~=~$10^{5}$~kg/m\textsuperscript{2} more volume segments ($3 \times 10^{5}$ and $6 \times 10^{5}$ respectively) were used for convergence as noted above. Note that the difference between the 32~Myr here for large $k$ values and the 26~Myr for the EBY11 model is due to the change from an equilibrium temperature of 25~K for the EBY11 model to our nominal equilibrium temperature of 250K.

\subsection{Kinetic Energy Imparted by Re-impacting Debris}
\label{sec:Results:KE}

\begin{figure}[ht]
\includegraphics[width=0.75\textwidth]{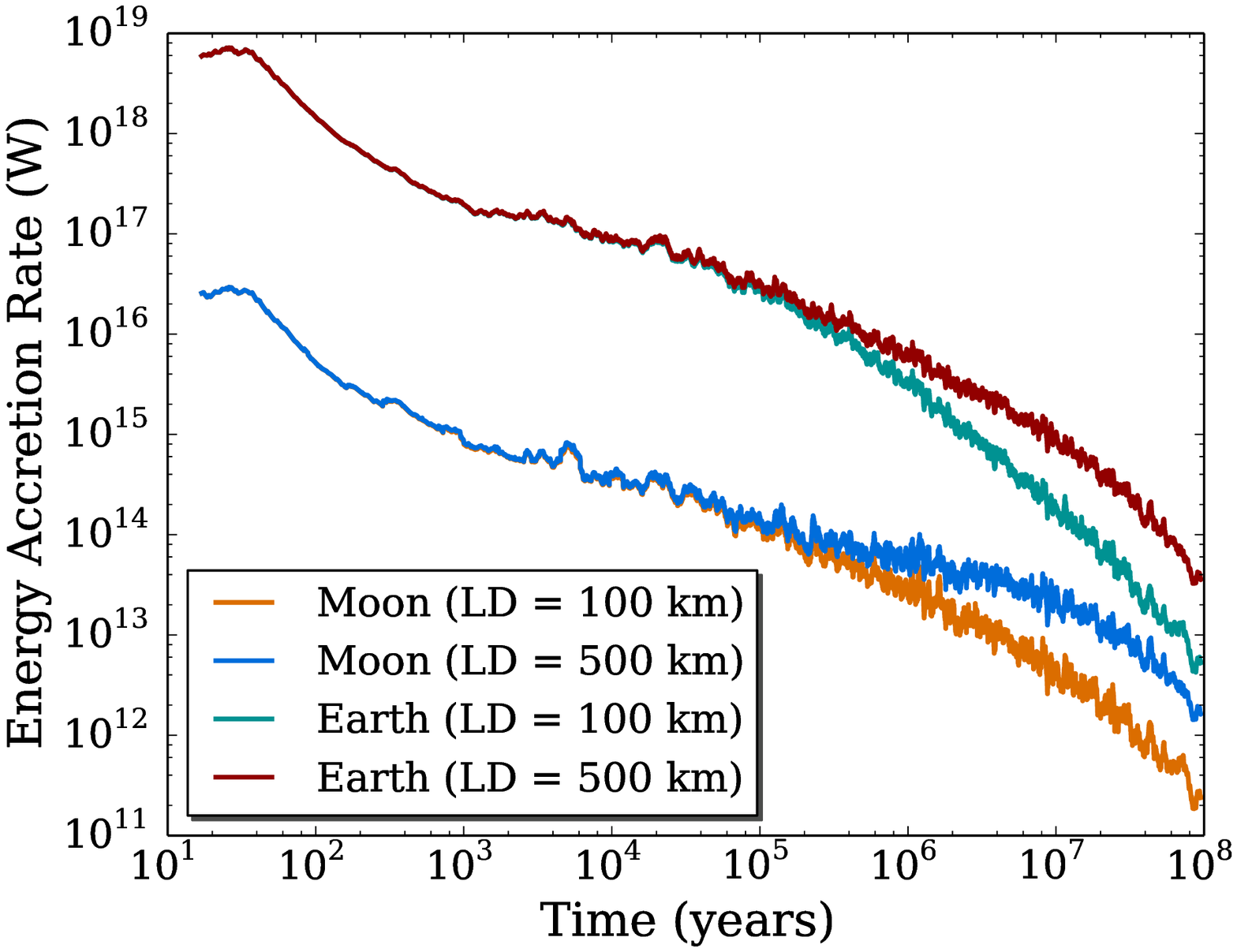}
\centering
\caption{\textcolor{black}{Energy accretion rate over time for the Earth (dark cyan and red lines) and the Moon (blue and orange lines) for two populations of re-impacting debris based on the size of the largest debris (LD). The Moon is at a distance of 10 Earth radii.}}
\label{fig:DebrisEnergy}
\end{figure}

\begin{figure}
\includegraphics[width=0.75\textwidth]{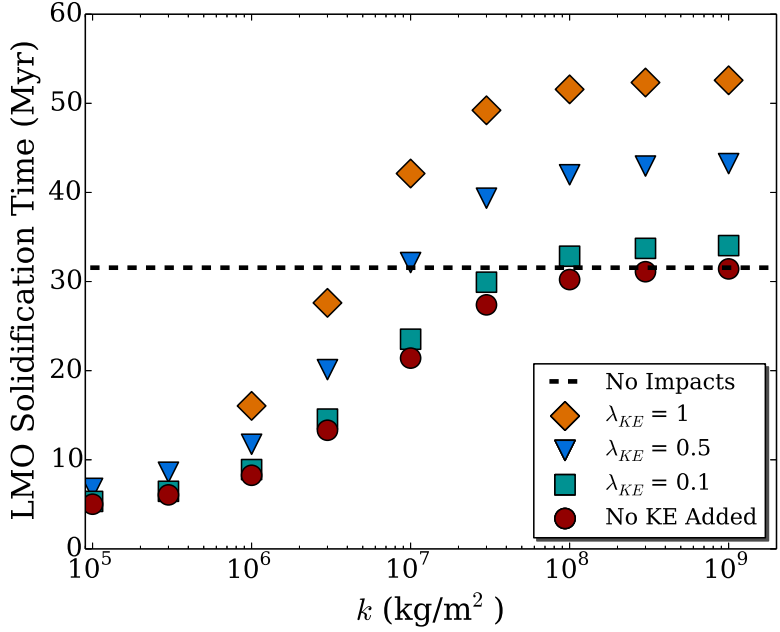}
\centering
\caption{LMO solidification time as a function of $k$ for different assumptions regarding kinetic energy imparted by re-impacting debris. $\lambda_{KE}$, kinetic energy efficiency, of 1 signifies that all of the kinetic energy of the impactors was imparted as thermal energy. The no impacts solidification time is shown with a dashed black line for reference.}
\label{fig:ImpartedKE}
\end{figure}

As mentioned in Section \ref{sec:Methods:Debris}, re-impacting debris will not only produce holes in the lunar crust but they will also impart thermal energy due to their kinetic energy (see Figure \ref{fig:DebrisEnergy}). Thus far we have only considered the effect of re-impacting debris producing holes. In Figure \ref{fig:ImpartedKE} we show the variation of the LMO solidification time as a result of different assumptions regarding the efficiency by which the impactors' kinetic energy is converted to thermal energy. We define that efficiency, $\lambda_{KE}$, to range from 1 (all of the kinetic energy is converted to thermal energy) to 0 (none of the kinetic energy is converted to thermal energy). \textcolor{black}{Note that in reality $\lambda_{KE}$ should never be exactly 1 since some energy is required to open the hole itself, however it is unclear how much of the kinetic energy would be required for hole opening and so we use 0 and 1 as the extreme possibilities.}

When kinetic energy is not considered, the effect of re-impacting debris is to reduce the LMO solidification time from $\sim$~32~Myr ($k = 10^{9}$~kg/m\textsuperscript{2}) to $\sim$~5~Myr ($k = 10^{5}$~kg/m\textsuperscript{2}). Interestingly, when we consider both hole production and thermal energy impartment by re-impacting debris, the LMO solidification time may be longer or shorter than the no impacts solidification time (i.e. $\sim$~32~Myr). If fewer holes are produced (i.e. $k > 10^{7}$~kg/m\textsuperscript{2}) and impacts are efficient at delivering thermal energy (i.e. $\lambda_{KE} > 0.5$), the LMO solidification time is greater than its value when impacts are not considered. On the other hand, regardless of $\lambda_{KE}$, if impacts generate a larger number of holes (i.e. $k < 3 \times 10^{6}$~kg/m\textsuperscript{2}), the LMO solidification time is less than its value when impacts are not considered. Thus, there are particular values of $k$ and $\lambda_{KE}$ that balance the increased amount of heat out due to holes and the additional heat input due to the kinetic energy of impacts. This would mean that for those values of $k$ and $\lambda_{KE}$, the LMO solidification time would be the same, with or without impacts.

The number of volume segments used for the different values of $k$ are the same as mentioned early. However, for $\lambda_{KE} = 1$, $k$~=~$3 \times 10^{8}$~kg/m\textsuperscript{2} required $5 \times 10^{5}$ volume segments and $k$~=~$10^{6}$~kg/m\textsuperscript{2} required $6 \times 10^{5}$ volume segments for convergence. For $\lambda_{KE} = 0.5$, $k$~=~$3 \times 10^{8}$~kg/m\textsuperscript{2} required $5 \times 10^{5}$ volume segments, $k$~=~$10^{5}$~kg/m\textsuperscript{2} required $7 \times 10^{5}$ volume segments, and $k$~=~$3 \times 10^{5}$~kg/m\textsuperscript{2} required $10^{6}$ volume segments. The larger number of segments required is a result of the increased amount of heat that must be lost from the LMO when impactor kinetic energy is considered.

\subsection{Concentrating Floatation Crust into Holes}
\label{sec:Results:Concentrating}

So far we evenly divided newly formed floatation crust between existing crust and holes according to the surface area covered by each. There \textcolor{black}{are} a number of reasons we might expect that more of the newly formed plagioclase would be drawn to the holes, however. Due to the increased thermal flux from the holes it is likely that convective upwellings would be located beneath them, which could concentrate new floatation crust into the holes.  Additionally, in much the same way as a ball will tend to roll down a hill, since the holes represent highs in the topography of the base of the crust bouyant material will tend to roll towards the holes.  \textcolor{black}{We can define the fraction of newly formed floatation crust that is directed into the holes as}
\begin{equation}
    C_{\rm holes} = f_e A_{\rm holes}/A_{\rm Moon},
\end{equation}
\textcolor{black}{where $A_{\rm holes}$ is the area covered by holes, $A_{\rm Moon}$ is the total surface area of the Moon and we introduce the enhancement factor, $f_e$.}
Our nominal case has an enhancement factor ($f_e$) of 1, such that equal amounts of new floatation crust per unit area go to both the holes and the rest of the crust, but we can vary $f_e$ to force more floatation crust into the holes.  \textcolor{black}{Note that $C_{\rm holes}$ is limited to be at most 1, since at that point all newly formed crust is being directed into the holes, and this effectively puts a limit on $f_e$, especially when the area covered by holes is large.} Using our nominal values with $k$~=~$10^7$~kg/m\textsuperscript{2} and $f_e$~=~1, the LMO solidification time is $\sim$~21~Myr. LMO solidification time increases to $\sim$~27~Myr and $\sim$~29~Myr for $f_e$~=~3 and $f_e$~=~10 respectively, with even higher $f_e$ values producing results that are very similar to the $f_e$~=~10 case.

Another outcome of interest is the surface area of the Moon that has holes at the end of the iteration. Holes are considered closed when their crustal thickness is equal to the non-impacted crustal thickness. Thus, in our nominal case, holes do not close since their crustal thicknesses will always be less than the non-impacted crustal thickness. If holes acquire an enhanced amount of floatation crust, they will close and thus reduce the area of the Moon that has holes at the end of the iteration. That area is 34\% for our nominal case, while it is 3\% and 0.7\% when $f_e$~=~3 and $f_e$~=~10 respectively. Though floatation crust enhancement has a significant effect on the surface area of holes; overall, it has little effect on the LMO solidification time.

% DISCUSSION
\section{Discussion}
\label{sec:Discussion}

\subsection{Reconciling Crust Sample Ages with the Magma Ocean Solidification Time}
\label{sec:Discussion:CrustAge} 

The LMO model predicts that the ages of the primordial lunar crust will be determined by the LMO solidification time. Assuming that lunar crustal ages have not been reset, the crust that formed from the floating anorthosite rocks should have an age that decreases with increasing depth below the surface. We would expect the age difference between the top and bottom layers of the crust should approximately be the time that it took for the last 20\% of the LMO to solidify. Age dating of lunar FAN samples have implied that the LMO may have taken over 200~Myr to solidify \citep{Alibert_1994, Borg_1999}. However, geochemical modeling work by \citet{ElkinsTanton_2011} showed that the LMO should have taken about 10~Myr to solidify, and we find a maximum solidification time of around 50 Myr with the probably somewhat unrealistic scenario of minimal hole puncturing and maximal impact energy deposition. Therefore, there is a discrepancy between the lunar crust sample ages and cooling models of the LMO. The reasons for this discrepancy may be due to one or both of the following: misinterpretation of lunar crust sample ages or the presence of additional heat sources for the LMO. We will discuss each of these in turn.

\subsubsection{\textcolor{black}{Lunar crustal age dating}}
\label{sec:discuss:crustage:samples}

It is possible that the primordial lunar crustal ages do not actually span 200~Myr. As discussed by \citet{Borg_2011}, some FAN samples, such as sample 60025 with an age of 4.360 $\pm$ 0.003 Gya, may have recorded more recent melting events rather than the formation time of the primordial lunar crust.  If this is the case, then samples with older ages such as samples 67016c (4.53 $\pm$ 0.12 Gya, \citealt{Shirley_1983}), 67075 (4.47 $\pm$ 0.07 Gya, \citealt{Nyquist_2010}), and Y-86032 (4.43 $\pm$ 0.03 Gya, \citealt{Nyquist_2006}) may have recorded the crystallization time of the primordial crust, while the younger samples may have recorded more recent re-melting events. These melting evens may be due to both `background' asteroidal impacts \citep{Nyquist_2006, Rolf_2017} and re-impacting debris \citep{Taylor_1993}. Evidence to the recrystallization of some parts of the lunar crust is given by both \citet{Ogawa_2011} and \citet{Yamamoto_2015b} who argue that high-calcium pyroxene material near young lunar craters was due to re-differentiation of the primordial anorthosite crust due to impacts.

To help elucidate these ages, it may be possible to delineate ages by the crystallization time of ur-KREEP material, which is estimated to have taken place at 4.368 $\pm$ 0.029 Gya \citep{Gaffney_2014, Borg_2015}. If ur-KREEP is identified with the final, incompatible remnant dregs of the LMO, the ur-KREEP crystallization time would represent an upper limit to the LMO solidification time since it must have solidified after the rest of the LMO. Hence, if we assume that the LMO solidification time is given by the difference between the ages of the oldest crust samples and the time of ur-KREEP crystallization, the crust samples would then indicate that the LMO solidification time was less than 160~Myr instead of 200~Myr. Taking the ur-KREEP crystallization time as an upper limit to the LMO solidification time improves the discrepancy between crustal age estimates and modelling, though it does not remove it completely.  That the ur-KREEP crystallization time is itself discrepant with some of the measured crustal ages is however suggestive that some re-examination of crustal age measurements may be in order.

\subsubsection{\textcolor{black}{Additional heat sources to the LMO}}
\label{sec:discuss:crustage:addheat}

As mentioned in Section \ref{sec:Methods:Code}, additional heat sources would have likely influenced the thermal evolution of the LMO. To explore this effect on the LMO, our code allows for additional constant heating. We use that to estimate the effects of heat sources such as tidal and radiogenic heating. Such heating is often approximately constant for the time interval we are interested in \citep[e.g.][]{Meyer_2010}.

\begin{figure}
\includegraphics[width=0.75\textwidth]{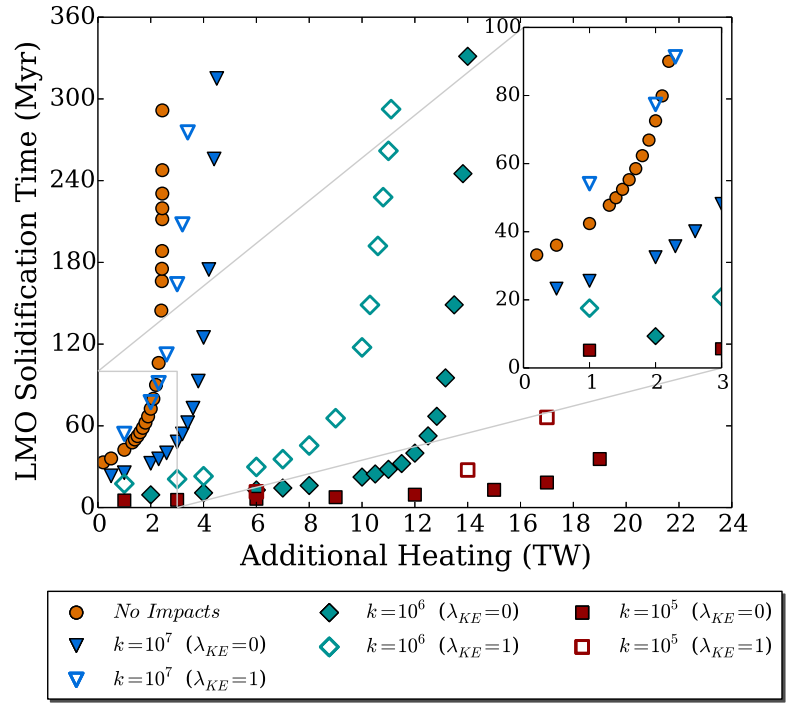}
\centering
\caption{\textcolor{black}{LMO solidification time as a function of additional constant heating. The without impacts case is shown with orange circles. The with impacts and with $\lambda_{KE}$~=~0 cases are shown with filled markers: blue triangles ($k$~=~$10^{7}$~kg/m\textsuperscript{2}), dark cyan diamonds ($k$~=~$10^{6}$~kg/m\textsuperscript{2}), and dark red squares ($k$~=~$10^{5}$~kg/m\textsuperscript{2}). The corresponding $\lambda_{KE}$~=~1 cases are shown with unfilled markers. The impact rate decays inversely with time beyond 100~Myr.}}
\label{fig:ExtraHeating}
\end{figure}

Using our nominal input parameter values, we varied this constant heating rate to see the result on the LMO solidification time. In Figure \ref{fig:ExtraHeating} we show the solidification time as a function of additional heating for the no impacts case and three impacts cases (with $k = 10^{5}$, $10^{6}$ \& $10^{7}$~kg/m\textsuperscript{2}). For the no impacts case, a heat rate of about $2.3\times 10^{12}$~W is sufficient to increase the LMO solidification time to about 200~Myrs. When impacts  are included, higher heat rates are required to increase the LMO solidification time to about 200~Myrs. For the case where $k = 10^7$~kg/m\textsuperscript{2}, the required heat rate is between 3.2 to $4.3\times 10^{12}$~W depending on $\lambda_{KE}$. For the case where $k\leq 10^6$~kg/m\textsuperscript{2} the required heat rate is more than $10^{13}$~W.  \textcolor{black}{Note that the response of the LMO solidification time to additional heating is non-linear, increasing rapidly as the additional heating approaches the maximum rate at which the magma ocean can lose heat.}

\textcolor{black}{We now want to compare the additional heating in Figure~\ref{fig:ExtraHeating} with the heating we expect from radiogenic heat sources, tides, and secular cooling of the core. The key radiogenic heat sources for the early Moon are $^{238}$U, $^{235}$U, $^{232}$Th and $^{40}$K.} \textcolor{black}{Since the Moon likely formed $\sim$~60~Myr after the formation of the solar system (with time-zero defined by the age of the oldest Ca-Al-rich inclusions [CAIs)]) \citep{Touboul_2009}, more energetic radiogenic heat sources such as $^{26}$Al would have been extinct due to their short half-lives.} \textcolor{black}{The abundance of uranium and thorium in the lunar mantle is estimated as being between 1 and 2 times that of Earth \citep[e.g.][and references therein]{Taylor_2006}, while potassium is depleted by a factor of around 4-8 \citep[e.g.][]{Oneill_1991, Taylor_2014b}. Extrapolating the present day abundances back 4.5~Gyr and using the heating rates from \citet{Turcotte_2014} results in a radiogenic heating rate of around 1.5--3~TW for the early Moon.  Unless hole generation is low (high $k$) and conversion of impact kinetic energy high this suggests that radiogenic heating alone is unlikely to be sufficient to extend the solidification time of the LMO to $\sim$200~Myr. Note that the shortest lived isotope ($^{235}$U) has a lifetime of 700~Myr, substantially longer than the timescales we are considering and so the fall in the radiogenic heating rate during the solidification of the LMO is small.} \textcolor{black}{As we can see from Figure~\ref{fig:ExtraHeating} the heating rates expected for radiogenic heating can only significantly influence the solidification time when $k$ is very large such that the effect of impacts is negligible, or moderate with very efficient kinetic energy conversion.  This suggests that radiogenic heating is unlikely to be sufficient to extend the solidification time of the LMO to $\sim$200~Myr.}

\textcolor{black}{Turning to tidal heating, }\citet{Meyer_2010} suggested that the typical tidal heating rate is around $6\times10^{11}$~W, lower than the roughly $10^{12}$~W they used for radiogenic heating.  \textcolor{black}{Note that the $\sim$200-300~Myr LMO solidification time obtained by \citet{Meyer_2010} is primarily due to their initially very hot and deep LMO (see \textcolor{black}{Appendix}~\ref{Apd:Sensitivity}), which we find unrealistic.}  This would not be sufficient to significantly increase the solidification time of the LMO if impact generated holes play an important role.  Calculations by \citet{Chen_2016} however suggest larger typical tidal heating rates of $\sim 4-8\times10^{12}$~W.  Tidal heating at these rates would be sufficient to substantially extend LMO solidification provided that hole production is moderate ($k>10^6$~kg/m\textsuperscript{2}).  Tidal heating rates can be much higher than these typical values when the lunar orbit passes through resonances \citep[e.g.][]{Touma_1998}, however these very high rates are short lived and occur early in the tidal evolution of the Moon.  Existing work thus suggests that tidal heating can only provide a large increase to the LMO solidification time if hole generation is not too vigorous.

\textcolor{black}{The remaining heating mechanism, secular cooling of the core, would not have been a likely candidate for sufficient heating since it's expected to have contributed $5 \times 10^{10}$~W of heating initially \citep{Konrad_1997}.}

%the other mechanism, electrical induction heating, may have been able to provide sufficient heating. Early work estimated that electrical induction heating to have provided $\sim$~$10^{14}$ to $10^{16}$~W of heating \citep{Sonett_1975}.

\textcolor{black}{It is also relevant to compare our additional heating rates in Figure~\ref{fig:ExtraHeating} with the impact energy accretion rates in Figure~\ref{fig:DebrisEnergy}. At earlier times the impact energy accretion rate is extremely high, exceeding $10^{16}$~W in the first 100~years, however it drops rapidly over time, falling to around 3~TW by $10^7$~years and around 0.1~TW by $10^8$~years. We can see this in Figure~\ref{fig:ExtraHeating} since the $\lambda_{KE}=1$ cases deviate substantially from the $\lambda_{KE}=1$ over the first few tens of Myr, but then run close to parallel.}

When considering the lunar crust sample ages, a caveat is that those samples are likely from the upper layers of the lunar crust. This means that we may be measuring time periods that only partially cover the time that it took for the LMO to solidify. It may be possible to obtain crust samples from the bottom of the crust by sampling certain impact craters, such as Moscoviense and Crisium, which may have been excavated down to the mantle \citep{Wieczorek_2013}. However, such samples are unlikely to solve the discrepancy between the LMO solidification times estimated by sample ages and those estimated by modeling work. A partial sampling of the crust would mean that the estimates based on crust samples, which are already much longer than estimates from modeling, should be even longer.

\subsection{Implications for the Lunar Surface}
\label{sec:Discussion:Surface} 
Re-impacting debris may have affected the primordial lunar crust by shock heating it, by puncturing holes into it, and by adding material onto it. While some of the changes to the crust caused by re-impacting debris may not be detectable today (such as crustal thickness variations due to refilled holes), there may be other effects that are recognizable. One potential effect, resetting of crustal ages by impacts, has already been mentioned in the previous section (Section \ref{sec:Discussion:CrustAge}). In the same manner, re-impacting debris may have produced feldspathic granulitic impactites \citep{Simonds_1974, Warner_1977} and granulitic noritic anorthosites \citep{McLeod_2016}. Re-impacting debris may have also aided the enrichment of \textsuperscript{37}Cl in lunar samples due to the degassing of \textsuperscript{35}Cl \citep{Sharp_2010, Boyce_2015} by breaching the primordial lunar crust \citep{Barnes_2016b}. Another consequence of breaching the crust is that quench crust would form in newly created holes. If the lunar surface was saturated with holes (i.e. $k \leq 10^{6}$~kg/m\textsuperscript{2}), at the completion of the LMO solidification approximately $\sim$~$1 \times 10^{18}$~kg (0.02\% of the lunar crust by mass) of quench crust would have formed and been incorporated into the primordial lunar crust. Lastly, re-impacting debris would have added material onto the lunar crust. As such, the early lunar crust should have comprised of largely anorthosite rock with some debris component, which had a similar composition to the original debris of the Moon-forming impact. We may assume that the LMO solidified in 60~Myr based on the average age differences between four samples (i.e. FAN sample 60025, norite samples 77215 and 78236/8, and troctolite sample 76535) that met all five of the reliability criteria identified in \citet{Borg_2015}. In that case, $\sim$~$2 \times 10^{18}$~kg of re-impacting debris (0.05\% of the lunar crust by mass) would be added onto the crust after the solidification of the LMO, which would be higher if the LMO solidified faster and if some fraction of impacting material does not penetrate the crust, but remains on the surface.

\subsection{Implications for the Lunar Interior and Orbital Evolution}
\label{sec:Discussion:Interior}

Re-impacting debris that punctured holes in the crust would have made the LMO an open magmatic system \citep[e.g.][]{O'Hara_1977} and thus may have altered its geochemical evolution. Periodically adding material that has the composition of the initial debris created during the Moon-forming impact may have altered the fractional crystallization process of the LMO. In our work, $\sim$~$2 \times 10^{20}$~kg of re-impacting debris would be added to the LMO. While this is a small fraction (0.33\%) of the initial LMO mass, addition of external material would become more significant as the LMO mass decreases. Thus, future geochemical modeling should allow for open system behavior to consider what effect it may have, both in the addition of material from debris and in loss of volatiles through degassing. One possible result of this may be the considerable variability of zinc in lunar anorthosite samples \citep{Kato_2015}. Another may be the heterogeneous distribution of water in the lunar interior. If debris stored and periodically added water to, or allowed it to be selectively degassed from, hole regions it may explain the contradiction between works claiming the Moon to be hydrous \citep{Saal_2008, Boyce_2010, McCubbin_2010} and works claiming it to be anhydrous \citep{Taylor_2006b, Sharp_2010}. Importantly, water content also controls the stability of plagioclase, with more water delaying plagioclase formation \citep{Lin_2017a}.

As discussed in Section \ref{sec:Discussion:CrustAge}, tidal heating is not only a possible external heat source for the LMO, but it may also be required to explain the range of lunar crust sample ages. Tidal heating is largely dependent on the eccentricity and the semi-major axis of the lunar orbit. Nonetheless, \citet{Tian_2017} showed that the Moon's initial tidal quality factor, $Q$, and rigidity along with how those values change over time are important to its orbital evolution. Both $Q$ and rigidity are linked to the internal structure of the moon, particularly the fraction that is liquid. As such the thermal evolution and the orbital evolution of the Moon are coupled. Previous work such as \citet{Meyer_2010} and \citet{Chen_2016} considered this coupling but they did not include the effect of re-impacting debris. Tides may also affect the holes themselves by producing cracks along lines of maximum stress. This should be more pronounced given that these surfaces are already weakened by impacts. Thus, tidal stress may make holes larger and prolong their closure, which can be explored in the future using finite element modeling.

% CONCLUSIONS
\section{Conclusions}
\label{sec:Conclusions}

The Moon likely formed after a giant impact. A large quantity of debris from that impact escaped the Earth-Moon system and subsequently returned over a period of 100~Myr. During that time, the Lunar Magma Ocean (LMO) would have been solidifying with an early quench crust, followed by an anorthositic crust on its surface. Re-impacting debris would have affected the thermal evolution of the LMO by puncturing holes into the crust and delivering thermal energy to the LMO. Holes that were produced would have increased the thermal flux that was initially limited by the conductive crust. While that would have sped up the solidification of the LMO, thermal energy imparted by impacts would have done the reverse, to an extent that is not yet clear. By investigating a wide range of possible values for the amount of hole generation and the efficiency of kinetic energy conversion by impacts, we suggest LMO solidification times ranging from $\sim$~5~Myr to $\sim$~50~Myr at the extrema. Given that the range of lunar crust sample ages may be 60 to 200~Myr, our lower estimates for the LMO solidification time would require one or more additional heat sources (e.g. tidal heating), potentially with very high heating rates to make the LMO solidification time consistent with the range of lunar crust sample ages. At the higher end, with moderate hole generation and efficient kinetic energy deposition, our work suggests that the amount of tidal heating required to bring the LMO solidification time into concordance with lunar crust sample ages may be less than previously thought, especially if the age span of samples that truly date LMO solidification is closer to 60~Myr rather than 200~Myr. With our simple model we have provided insight into how re-impacting debris influences the cooling time of the LMO.  Nonetheless, further work is still needed to integrate all aspects of the early thermal evolution of the Moon, including geochemistry, re-impacting debris, tides and crustal structure.

% ACKNOWLEDGMENTS
\acknowledgments
This work was supported by NASA grant NNX16AI31G. We would like to thank Mark Robinson for his insightful suggestions that helped us consider quench crust at the early stage of the lunar thermal evolution and Julian Lowman for helpful discussions regarding convection in the LMO. We would also like to thank Francis Nimmo for his thoughtful recommendations. \textcolor{black}{We appreciate the helpful suggestions provided by the anonymous reviewer and reviewer Matthieu Laneuville. The Python code developed in this work is available at https://github.com/virangaperera/iMagma under a GNU General Public License v3.0. Data generated in this work is available \\
at http://doi.org/10.5281/zenodo.1167860.}

\bibliography{LMOcool} 

\begin{thebibliography}{133}
\providecommand{\natexlab}[1]{#1}
\expandafter\ifx\csname urlstyle\endcsname\relax
  \providecommand{\doi}[1]{doi:\discretionary{}{}{}#1}\else
  \providecommand{\doi}{doi:\discretionary{}{}{}\begingroup
  \urlstyle{rm}\Url}\fi

\bibitem[{\textit{Abe and Matsui}(1986)}]{Abe_1986}
Abe, Y., and T.~Matsui (1986), Early evolution of the earth: Accretion,
  atmosphere formation, and thermal history, \textit{Journal of Geophysical
  Research: Solid Earth}, \textit{91}(B13), E291--E302,
  \doi{10.1029/JB091iB13p0E291}.

\bibitem[{\textit{Alibert et~al.}(1994)\textit{Alibert, Norman, and
  McCulloch}}]{Alibert_1994}
Alibert, C., M.~D. Norman, and M.~T. McCulloch (1994), {An ancient Sm-Nd age
  for a ferroan noritic anorthosite clast from lunar breccia 67016},
  \textit{Geochimica et Cosmochimica Acta}, \textit{58}(13), 2921 -- 2926,
  \doi{10.1016/0016-7037(94)90125-2}.

\bibitem[{\textit{Andrews-Hanna et~al.}(2013)\textit{Andrews-Hanna, Asmar,
  Head, Kiefer, Konopliv, Lemoine, Matsuyama, Mazarico, McGovern, Melosh,
  Neumann, Nimmo, Phillips, Smith, Solomon, Taylor, Wieczorek, Williams, and
  Zuber}}]{AndrewsHanna_2013}
Andrews-Hanna, J.~C., S.~W. Asmar, J.~W. Head, W.~S. Kiefer, A.~S. Konopliv,
  F.~G. Lemoine, I.~Matsuyama, E.~Mazarico, P.~J. McGovern, H.~J. Melosh, G.~A.
  Neumann, F.~Nimmo, R.~J. Phillips, D.~E. Smith, S.~C. Solomon, G.~J. Taylor,
  M.~A. Wieczorek, J.~G. Williams, and M.~T. Zuber (2013), {Ancient Igneous
  Intrusions and Early Expansion of the Moon Revealed by GRAIL Gravity
  Gradiometry}, \textit{Science}, \textit{339}(6120), 675--678,
  \doi{10.1126/science.1231753}.

\bibitem[{\textit{Asphaug}(2014)}]{Asphaug_2014}
Asphaug, E. (2014), {Impact Origin of the Moon?}, \textit{Annual Review of
  Earth and Planetary Sciences}, \textit{42}(1), 551--578,
  \doi{10.1146/annurev-earth-050212-124057}.

\bibitem[{\textit{{Bandermann} and {Singer}}(1973)}]{Bandermann_1973}
{Bandermann}, L.~W., and S.~F. {Singer} (1973), {Calculation of Meteoroid
  Impacts on Moon and Earth}, \textit{Icarus}, \textit{19}, 108--113,
  \doi{10.1016/0019-1035(73)90142-5}.

\bibitem[{\textit{Barnes et~al.}(2016)\textit{Barnes, Tart\`{e}se, Anand,
  McCubbin, Neal, and Franchi}}]{Barnes_2016b}
Barnes, J.~J., R.~Tart\`{e}se, M.~Anand, F.~M. McCubbin, C.~R. Neal, and I.~A.
  Franchi (2016), {Early degassing of lunar urKREEP by crust-breaching
  impact(s)}, \textit{Earth and Planetary Science Letters}, \textit{447},
  84--94, \doi{10.1016/j.epsl.2016.04.036}.

\bibitem[{\textit{Barr}(2016)}]{Barr_2016}
Barr, A.~C. (2016), {On the origin of Earth's Moon}, \textit{Journal of
  Geophysical Research: Planets}, \textit{121}(9), 1573--1601,
  \doi{10.1002/2016JE005098}.

\bibitem[{\textit{{Bauer} and {Cox}}(2011)}]{Bauer_2011}
{Bauer}, A.~W., and R.~{Cox} (2011), {Hydrocode Modeling of Impacts at Europa},
  in \textit{Lunar and Planetary Science Conference}, vol.~42, p. 1123.

\bibitem[{\textit{Borg et~al.}(1999)\textit{Borg, Norman, Nyquist, Bogard,
  Snyder, Taylor, and Lindstrom}}]{Borg_1999}
Borg, L., M.~Norman, L.~Nyquist, D.~Bogard, G.~Snyder, L.~Taylor, and
  M.~Lindstrom (1999), {Isotopic studies of ferroan anorthosite 62236: a young
  lunar crustal rock from a light rare-earth-element-depleted source},
  \textit{Geochimica et Cosmochimica Acta}, \textit{63}(17), 2679 -- 2691,
  \doi{10.1016/S0016-7037(99)00130-1}.

\bibitem[{\textit{{Borg} et~al.}(2011)\textit{{Borg}, {Connelly}, {Boyet}, and
  {Carlson}}}]{Borg_2011}
{Borg}, L.~E., J.~N. {Connelly}, M.~{Boyet}, and R.~W. {Carlson} (2011),
  {Chronological evidence that the Moon is either young or did not have a
  global magma ocean}, \textit{Nature}, \textit{477}, 70--72,
  \doi{10.1038/nature10328}.

\bibitem[{\textit{Borg et~al.}(2015)\textit{Borg, Gaffney, and
  Shearer}}]{Borg_2015}
Borg, L.~E., A.~M. Gaffney, and C.~K. Shearer (2015), {A review of lunar
  chronology revealing a preponderance of 4.34--4.37 Ga ages},
  \textit{Meteoritics \& Planetary Science}, \textit{50}(4), 715--732,
  \doi{10.1111/maps.12373}.

\bibitem[{\textit{Bottinga and Weill}(1972)}]{Bottinga_1972}
Bottinga, Y., and D.~F. Weill (1972), {The viscosity of magmatic silicate
  liquids; a model calculation}, \textit{American Journal of Science},
  \textit{272}(5), 438--475, \doi{10.2475/ajs.272.5.438}.

\bibitem[{\textit{{Boyce} et~al.}(2010)\textit{{Boyce}, {Liu}, {Rossman},
  {Guan}, {Eiler}, {Stolper}, and {Taylor}}}]{Boyce_2010}
{Boyce}, J.~W., Y.~{Liu}, G.~R. {Rossman}, Y.~{Guan}, J.~M. {Eiler}, E.~M.
  {Stolper}, and L.~A. {Taylor} (2010), {Lunar apatite with terrestrial
  volatile abundances}, \textit{Nature}, \textit{466}, 466--469,
  \doi{10.1038/nature09274}.

\bibitem[{\textit{Boyce et~al.}(2015)\textit{Boyce, Treiman, Guan, Ma, Eiler,
  Gross, Greenwood, and Stolper}}]{Boyce_2015}
Boyce, J.~W., A.~H. Treiman, Y.~Guan, C.~Ma, J.~M. Eiler, J.~Gross, J.~P.
  Greenwood, and E.~M. Stolper (2015), {The chlorine isotope fingerprint of the
  lunar magma ocean}, \textit{Science Advances}, \textit{1}(8),
  \doi{10.1126/sciadv.1500380}.

\bibitem[{\textit{Boyet and Carlson}(2007)}]{Boyet_2007}
Boyet, M., and R.~W. Carlson (2007), {A highly depleted moon or a non-magma
  ocean origin for the lunar crust?}, \textit{Earth and Planetary Science
  Letters}, \textit{262}(3--4), 505--516, \doi{10.1016/j.epsl.2007.08.009}.

\bibitem[{\textit{Breuer and Moore}(2015)}]{Breuer_2015}
Breuer, D., and W.~Moore (2015), {Dynamics and Thermal History of the
  Terrestrial Planets, the Moon, and Io}, in \textit{Treatise on Geophysics},
  edited by G.~Schubert, 2nd ed., pp. 255 -- 305, Elsevier, Oxford,
  \doi{10.1016/B978-0-444-53802-4.00173-1}.

\bibitem[{\textit{{Cameron} and {Ward}}(1976)}]{Cameron_1976}
{Cameron}, A.~G.~W., and W.~R. {Ward} (1976), {The Origin of the Moon}, in
  \textit{Lunar and Planetary Inst.~Technical Report}, vol.~7.

\bibitem[{\textit{{Canup}}(2004)}]{Canup_2004}
{Canup}, R.~M. (2004), {Dynamics of Lunar Formation}, \textit{Annu. Rev.
  Astron. Astrophys.}, \textit{42}, 441--475,
  \doi{10.1146/annurev.astro.41.082201.113457}.

\bibitem[{\textit{Canup}(2012)}]{Canup_2012}
Canup, R.~M. (2012), {Forming a Moon with an Earth-like Composition via a Giant
  Impact}, \textit{Science}, \textit{338}(6110), 1052--1055,
  \doi{10.1126/science.1226073}.

\bibitem[{\textit{{Chambers}}(1999)}]{Chambers_1999}
{Chambers}, J.~E. (1999), {A hybrid symplectic integrator that permits close
  encounters between massive bodies}, \textit{Monthly Notices of the Royal
  Astronomical Society}, \textit{304}, 793--799,
  \doi{10.1046/j.1365-8711.1999.02379.x}.

\bibitem[{\textit{Chen and Nimmo}(2016)}]{Chen_2016}
Chen, E.~M., and F.~Nimmo (2016), {Tidal dissipation in the lunar magma ocean
  and its effect on the early evolution of the Earth--Moon system},
  \textit{Icarus}, \textit{275}, 132--142, \doi{10.1016/j.icarus.2016.04.012}.

\bibitem[{\textit{{\'C}uk and Stewart}(2012)}]{Cuk_2012}
{\'C}uk, M., and S.~T. Stewart (2012), {Making the Moon from a Fast-Spinning
  Earth: A Giant Impact Followed by Resonant Despinning}, \textit{Science},
  \textit{338}(6110), 1047--1052, \doi{10.1126/science.1225542}.

\bibitem[{\textit{{{\'C}uk} et~al.}(2016)\textit{{{\'C}uk}, {Hamilton}, {Lock},
  and {Stewart}}}]{Cuk_2016}
{{\'C}uk}, M., D.~P. {Hamilton}, S.~J. {Lock}, and S.~T. {Stewart} (2016),
  {Tidal evolution of the Moon from a high-obliquity, high-angular-momentum
  Earth}, \textit{Nature}, \textit{539}, 402--406, \doi{10.1038/nature19846}.

\bibitem[{\textit{Daly}(1946)}]{Daly_1946}
Daly, R.~A. (1946), {Origin of the Moon and Its Topography},
  \textit{Proceedings of the American Philosophical Society}, \textit{90}(2),
  104--119.

\bibitem[{\textit{{Davies}}(1982)}]{Davies_1982}
{Davies}, G.~F. (1982), {Impact Disruption of Magma Ocean Crust}, in
  \textit{Lunar and Planetary Science Conference}, vol.~13, p. 141.

\bibitem[{\textit{{Dohnanyi}}(1969)}]{Dohnanyi_1969}
{Dohnanyi}, J.~S. (1969), {Collisional Model of Asteroids and Their Debris},
  \textit{J. Geophys. Res.}, \textit{74}, 2531--2554,
  \doi{10.1029/JB074i010p02531}.

\bibitem[{\textit{Elkins-Tanton}(2012)}]{ElkinsTanton_2012}
Elkins-Tanton, L.~T. (2012), {Magma Oceans in the Inner Solar System},
  \textit{Annual Review of Earth and Planetary Sciences}, \textit{40}(1),
  113--139, \doi{10.1146/annurev-earth-042711-105503}.

\bibitem[{\textit{Elkins-Tanton et~al.}(2007)\textit{Elkins-Tanton, Smrekar,
  Hess, and Parmentier}}]{ElkinsTanton_2007}
Elkins-Tanton, L.~T., S.~E. Smrekar, P.~C. Hess, and E.~M. Parmentier (2007),
  {Volcanism and volatile recycling on a one-plate planet: Applications to
  Venus}, \textit{Journal of Geophysical Research: Planets}, \textit{112}(E4),
  \doi{10.1029/2006JE002793}.

\bibitem[{\textit{Elkins-Tanton et~al.}(2011)\textit{Elkins-Tanton, Burgess,
  and Yin}}]{ElkinsTanton_2011}
Elkins-Tanton, L.~T., S.~Burgess, and Q.-Z. Yin (2011), {The lunar magma ocean:
  Reconciling the solidification process with lunar petrology and
  geochronology}, \textit{Earth and Planetary Science Letters},
  \textit{304}(3--4), 326 -- 336, \doi{{10.1016/j.epsl.2011.02.004}}.

\bibitem[{\textit{Eppelbaum et~al.}(2014)\textit{Eppelbaum, Kutasov, and
  Pilchin}}]{Eppelbaum_2014}
Eppelbaum, L., I.~Kutasov, and A.~Pilchin (2014), \textit{{Thermal Properties
  of Rocks and Density of Fluids}}, pp. 99--149, Springer Berlin Heidelberg,
  Berlin, Heidelberg, \doi{10.1007/978-3-642-34023-9\_2}.

\bibitem[{\textit{Gaffney and Borg}(2014)}]{Gaffney_2014}
Gaffney, A.~M., and L.~E. Borg (2014), {A young solidification age for the
  lunar magma ocean}, \textit{Geochimica et Cosmochimica Acta}, \textit{140},
  227--240, \doi{10.1016/j.gca.2014.05.028}.

\bibitem[{\textit{Gast and Giuli}(1972)}]{Gast_1972b}
Gast, P.~W., and R.~T. Giuli (1972), {Density of the lunar interior},
  \textit{Earth and Planetary Science Letters}, \textit{16}(2), 299 -- 305,
  \doi{10.1016/0012-821X(72)90206-3}.

\bibitem[{\textit{Hartmann}(1980)}]{Hartmann_1980}
Hartmann, W. (1980), {Dropping stones in magma oceans: Effects of early lunar
  cratering}, in \textit{Lunar highlands crust}, pp. 155--171.

\bibitem[{\textit{Hartmann and Davis}(1975)}]{Hartmann_1975}
Hartmann, W.~K., and D.~R. Davis (1975), {Satellite-sized planetesimals and
  lunar origin}, \textit{Icarus}, \textit{24}(4), 504 -- 515,
  \doi{10.1016/0019-1035(75)90070-6}.

\bibitem[{\textit{{Hodges} and {Kushiro}}(1974)}]{Hodges_1974}
{Hodges}, F.~N., and I.~{Kushiro} (1974), {Apollo 17 petrology and experimental
  determination of differentiation sequences in model moon compositions}, in
  \textit{Lunar and Planetary Science Conference Proceedings}, vol.~5, pp.
  505--520.

\bibitem[{\textit{{Ida} et~al.}(1997)\textit{{Ida}, {Canup}, and
  {Stewart}}}]{Ida_1997}
{Ida}, S., R.~M. {Canup}, and G.~R. {Stewart} (1997), {Lunar accretion from an
  impact-generated disk}, \textit{Nature}, \textit{389}, 353--357,
  \doi{10.1038/38669}.

\bibitem[{\textit{Jackson and Wyatt}(2012)}]{Jackson_2012}
Jackson, A.~P., and M.~C. Wyatt (2012), {Debris from terrestrial planet
  formation: the Moon-forming collision}, \textit{Monthly Notices of the Royal
  Astronomical Society}, \textit{425}(1), 657--679,
  \doi{10.1111/j.1365-2966.2012.21546.x}.

\bibitem[{\textit{{Jackson} et~al.}(2014)\textit{{Jackson}, {Wyatt}, {Bonsor},
  and {Veras}}}]{Jackson_2014}
{Jackson}, A.~P., M.~C. {Wyatt}, A.~{Bonsor}, and D.~{Veras} (2014), {Debris
  froms giant impacts between planetary embryos at large orbital radii},
  \textit{Monthly Notices of the Royal Astronomical Society}, \textit{440},
  3757--3777, \doi{10.1093/mnras/stu476}.

\bibitem[{\textit{{Kato} et~al.}(2015)\textit{{Kato}, {Moynier}, {Valdes},
  {Dhaliwal}, and {Day}}}]{Kato_2015}
{Kato}, C., F.~{Moynier}, M.~C. {Valdes}, J.~K. {Dhaliwal}, and J.~M.~D. {Day}
  (2015), {Extensive volatile loss during formation and differentiation of the
  Moon}, \textit{Nature Communications}, \textit{6}, 7617,
  \doi{10.1038/ncomms8617}.

\bibitem[{\textit{Kaula}(1979)}]{Kaula_1979}
Kaula, W.~M. (1979), {Thermal evolution of Earth and Moon growing by
  planetesimal impacts}, \textit{Journal of Geophysical Research: Solid Earth},
  \textit{84}(B3), 999--1008, \doi{10.1029/JB084iB03p00999}.

\bibitem[{\textit{Kirk and Stevenson}(1989)}]{Kirk_1989}
Kirk, R.~L., and D.~J. Stevenson (1989), {The competition between thermal
  contraction and differentiation in the stress history of the Moon},
  \textit{Journal of Geophysical Research: Solid Earth}, \textit{94}(B9),
  12,133--12,144, \doi{10.1029/JB094iB09p12133}.

\bibitem[{\textit{Kokubo et~al.}(2000{\natexlab{a}})\textit{Kokubo, Ida, and
  Makino}}]{Kokubo_2000}
Kokubo, E., S.~Ida, and J.~Makino (2000{\natexlab{a}}), {Evolution of a
  Circumterrestrial Disk and Formation of a Single Moon}, \textit{Icarus},
  \textit{148}(2), 419--436, \doi{10.1006/icar.2000.6496}.

\bibitem[{\textit{Kokubo et~al.}(2000{\natexlab{b}})\textit{Kokubo, Canup, and
  Ida}}]{Kokubo_2000b}
Kokubo, E., R.~Canup, and S.~Ida (2000{\natexlab{b}}), {Lunar accretion from an
  impact-generated disk}, \textit{Origin of the Earth and Moon}, pp. 145--163.

\bibitem[{\textit{{Konrad} and {Spohn}}(1997)}]{Konrad_1997}
{Konrad}, W., and T.~{Spohn} (1997), {Thermal history of the moon--Implications
  for an early core dynamo and post-accretional magmatism}, \textit{Advances in
  Space Research}, \textit{19}, 1511, \doi{10.1016/S0273-1177(97)00364-5}.

\bibitem[{\textit{{Leinhardt} and {Stewart}}(2012)}]{Leinhardt_2012}
{Leinhardt}, Z.~M., and S.~T. {Stewart} (2012), {Collisions between
  Gravity-dominated Bodies. I. Outcome Regimes and Scaling Laws},
  \textit{Astrophys. J.}, \textit{745}, 79, \doi{10.1088/0004-637X/745/1/79}.

\bibitem[{\textit{{Lin} et~al.}(2017{\natexlab{a}})\textit{{Lin}, {Tronche},
  {Steenstra}, and {van Westrenen}}}]{Lin_2017b}
{Lin}, Y., E.~J. {Tronche}, E.~S. {Steenstra}, and W.~{van Westrenen}
  (2017{\natexlab{a}}), {Experimental constraints on the solidification of a
  nominally dry lunar magma ocean}, \textit{Earth and Planetary Science
  Letters}, \textit{471}, 104 -- 116, \doi{10.1016/j.epsl.2017.04.045}.

\bibitem[{\textit{{Lin} et~al.}(2017{\natexlab{b}})\textit{{Lin}, {Tronche},
  {Steenstra}, and {van Westrenen}}}]{Lin_2017a}
{Lin}, Y., E.~J. {Tronche}, E.~S. {Steenstra}, and W.~{van Westrenen}
  (2017{\natexlab{b}}), {Evidence for an early wet Moon from experimental
  crystallization of the lunar magma ocean}, \textit{Nature Geoscience},
  \textit{10}, 14--18, \doi{10.1038/ngeo2845}.

\bibitem[{\textit{Longhi}(1980)}]{Longhi_1980}
Longhi, J. (1980), {A model of early lunar differentiation}, in \textit{Lunar
  and Planetary Science Conference Proceedings}, vol.~11, pp. 289--315.

\bibitem[{\textit{Longhi}(2003)}]{Longhi_2003}
Longhi, J. (2003), {A new view of lunar ferroan anorthosites: Postmagma ocean
  petrogenesis}, \textit{Journal of Geophysical Research: Planets},
  \textit{108}(E8), \doi{10.1029/2002JE001941}.

\bibitem[{\textit{Longhi and Ashwal}(1985)}]{Longhi_1985}
Longhi, J., and L.~D. Ashwal (1985), {Two-stage models for lunar and
  terrestrial anorthosites: Petrogenesis without a magma ocean},
  \textit{Journal of Geophysical Research: Solid Earth}, \textit{90}(S02),
  C571--C584, \doi{10.1029/JB090iS02p0C571}.

\bibitem[{\textit{Luhmann et~al.}(1992)\textit{Luhmann, Johnson, and
  Zhang}}]{Luhmann_1992}
Luhmann, J.~G., R.~E. Johnson, and M.~H.~G. Zhang (1992), {Evolutionary impact
  of sputtering of the Martian atmosphere by O+ pickup ions},
  \textit{Geophysical Research Letters}, \textit{19}(21), 2151--2154,
  \doi{10.1029/92GL02485}.

\bibitem[{\textit{Marcus et~al.}(2009)\textit{Marcus, Stewart, Sasselov, and
  Hernquist}}]{Marcus_2009}
Marcus, R.~A., S.~T. Stewart, D.~Sasselov, and L.~Hernquist (2009),
  {Collisional Stripping and Disruption of Super-Earths}, \textit{The
  Astrophysical Journal Letters}, \textit{700}(2), L118.

\bibitem[{\textit{{Mastrobuono-Battisti}
  et~al.}(2015)\textit{{Mastrobuono-Battisti}, {Perets}, and
  {Raymond}}}]{MastrobuonoBattisti_2015}
{Mastrobuono-Battisti}, A., H.~B. {Perets}, and S.~N. {Raymond} (2015), {A
  primordial origin for the compositional similarity between the Earth and the
  Moon}, \textit{Nature}, \textit{520}, 212--215, \doi{10.1038/nature14333}.

\bibitem[{\textit{Matson et~al.}(2006)\textit{Matson, Davies, Veeder, Rathbun,
  Johnson, and Castillo}}]{Matson_2006}
Matson, D.~L., A.~G. Davies, G.~J. Veeder, J.~A. Rathbun, T.~V. Johnson, and
  J.~C. Castillo (2006), {Io: Loki Patera as a magma sea}, \textit{Journal of
  Geophysical Research: Planets}, \textit{111}(E9), \doi{10.1029/2006JE002703}.

\bibitem[{\textit{Matsuyama et~al.}(2016)\textit{Matsuyama, Nimmo, Keane, Chan,
  Taylor, Wieczorek, Kiefer, and Williams}}]{Matsuyama_2016}
Matsuyama, I., F.~Nimmo, J.~T. Keane, N.~H. Chan, G.~J. Taylor, M.~A.
  Wieczorek, W.~S. Kiefer, and J.~G. Williams (2016), {GRAIL, LLR, and LOLA
  constraints on the interior structure of the Moon}, \textit{Geophysical
  Research Letters}, \textit{43}(16), 8365--8375, \doi{10.1002/2016GL069952}.

\bibitem[{\textit{{McCubbin} et~al.}(2010)\textit{{McCubbin}, {Steele},
  {Hauri}, {Nekvasil}, {Yamashita}, and {Hemley}}}]{McCubbin_2010}
{McCubbin}, F.~M., A.~{Steele}, E.~H. {Hauri}, H.~{Nekvasil}, S.~{Yamashita},
  and R.~J. {Hemley} (2010), {Nominally hydrous magmatism on the Moon},
  \textit{Proceedings of the National Academy of Science}, \textit{107},
  11,223--11,228, \doi{10.1073/pnas.1006677107}.

\bibitem[{\textit{McLeod et~al.}(2016)\textit{McLeod, Brandon, Fernandes,
  Peslier, Fritz, Lapen, Shafer, Butcher, and Irving}}]{McLeod_2016}
McLeod, C.~L., A.~D. Brandon, V.~A. Fernandes, A.~H. Peslier, J.~Fritz,
  T.~Lapen, J.~T. Shafer, A.~R. Butcher, and A.~J. Irving (2016), {Constraints
  on formation and evolution of the lunar crust from feldspathic granulitic
  breccias NWA 3163 and 4881}, \textit{Geochimica et Cosmochimica Acta},
  \textit{187}, 350--374, \doi{10.1016/j.gca.2016.04.032}.

\bibitem[{\textit{{Melosh} and {Vickery}}(1989)}]{Melosh_1989}
{Melosh}, H.~J., and A.~M. {Vickery} (1989), {Impact erosion of the primordial
  atmosphere of Mars}, \textit{Nature}, \textit{338}, 487--489,
  \doi{10.1038/338487a0}.

\bibitem[{\textit{Merk and Prialnik}(2006)}]{Merk_2006}
Merk, R., and D.~Prialnik (2006), {Combined modeling of thermal evolution and
  accretion of trans-neptunian objects--Occurrence of high temperatures and
  liquid water}, \textit{Icarus}, \textit{183}(2), 283 -- 295,
  \doi{10.1016/j.icarus.2006.02.011}.

\bibitem[{\textit{{Meyer} et~al.}(2010)\textit{{Meyer}, {Elkins-Tanton}, and
  {Wisdom}}}]{Meyer_2010}
{Meyer}, J., L.~{Elkins-Tanton}, and J.~{Wisdom} (2010), {Coupled
  thermal-orbital evolution of the early Moon}, \textit{Icarus}, \textit{208},
  1--10, \doi{10.1016/j.icarus.2010.01.029}.

\bibitem[{\textit{Minear and Fletcher}(1978)}]{Minear_1978}
Minear, J., and C.~Fletcher (1978), {Crystallization of a lunar magma ocean},
  in \textit{Lunar and Planetary Science Conference Proceedings}, vol.~9, pp.
  263--283.

\bibitem[{\textit{{Minear}}(1980)}]{Minear_1980}
{Minear}, J.~W. (1980), {The lunar magma ocean: A transient lunar phenomenon?},
  in \textit{Lunar and Planetary Science Conference Proceedings}, vol.~11,
  edited by S.~A. {Bedini}, pp. 1941--1955.

\bibitem[{\textit{Monteux et~al.}(2016)\textit{Monteux, Andrault, and
  Samuel}}]{Monteux_2016}
Monteux, J., D.~Andrault, and H.~Samuel (2016), {On the cooling of a deep
  terrestrial magma ocean}, \textit{Earth and Planetary Science Letters},
  \textit{448}, 140 -- 149, \doi{10.1016/j.epsl.2016.05.010}.

\bibitem[{\textit{{Nemchin} et~al.}(2009)\textit{{Nemchin}, {Timms}, {Pidgeon},
  {Geisler}, {Reddy}, and {Meyer}}}]{Nemchin_2009}
{Nemchin}, A., N.~{Timms}, R.~{Pidgeon}, T.~{Geisler}, S.~{Reddy}, and
  C.~{Meyer} (2009), {Timing of crystallization of the lunar magma ocean
  constrained by the oldest zircon}, \textit{Nature Geoscience}, \textit{2},
  133--136, \doi{10.1038/ngeo417}.

\bibitem[{\textit{Neumann et~al.}(2014)\textit{Neumann, Breuer, and
  Spohn}}]{Neumann_2014}
Neumann, W., D.~Breuer, and T.~Spohn (2014), {Differentiation of Vesta:
  Implications for a shallow magma ocean}, \textit{Earth and Planetary Science
  Letters}, \textit{395}, 267--280, \doi{10.1016/j.epsl.2014.03.033}.

\bibitem[{\textit{{Niemela} et~al.}(2000)\textit{{Niemela}, {Skrbek},
  {Sreenivasan}, and {Donnelly}}}]{Niemela_2000}
{Niemela}, J.~J., L.~{Skrbek}, K.~R. {Sreenivasan}, and R.~J. {Donnelly}
  (2000), {Turbulent convection at very high Rayleigh numbers},
  \textit{Nature}, \textit{404}, 837--840, \doi{10.1038/35009036}.

\bibitem[{\textit{Nyquist et~al.}(2006)\textit{Nyquist, Bogard, Yamaguchi,
  Shih, Karouji, Ebihara, Reese, Garrison, McKay, and Takeda}}]{Nyquist_2006}
Nyquist, L., D.~Bogard, A.~Yamaguchi, C.-Y. Shih, Y.~Karouji, M.~Ebihara,
  Y.~Reese, D.~Garrison, G.~McKay, and H.~Takeda (2006), {Feldspathic clasts in
  Yamato-86032: Remnants of the lunar crust with implications for its formation
  and impact history}, \textit{Geochimica et Cosmochimica Acta},
  \textit{70}(24), 5990--6015, \doi{10.1016/j.gca.2006.07.042}, a Special Issue
  Dedicated to Larry A. Haskin.

\bibitem[{\textit{{Nyquist} et~al.}(1995)\textit{{Nyquist}, {Wiesmann},
  {Bansal}, {Shih}, {Keith}, and {Harper}}}]{Nyquist_1995}
{Nyquist}, L.~E., H.~{Wiesmann}, B.~{Bansal}, C.-Y. {Shih}, J.~E. {Keith}, and
  C.~L. {Harper} (1995), {$^{146}$Sm-$^{142}$Nd formation interval for the
  lunar mantle}, \textit{Geochimica et Cosmochimica Acta}, \textit{59},
  2817--2837, \doi{10.1016/0016-7037(95)00175-Y}.

\bibitem[{\textit{{Nyquist} et~al.}(2010)\textit{{Nyquist}, {Shih}, {Reese},
  {Park}, {Bogard}, {Garrison}, and {Yamaguchi}}}]{Nyquist_2010}
{Nyquist}, L.~E., C.-Y. {Shih}, Y.~D. {Reese}, J.~{Park}, D.~D. {Bogard}, D.~H.
  {Garrison}, and A.~{Yamaguchi} (2010), {Lunar Crustal History Recorded in
  Lunar Anorthosites}, in \textit{Lunar and Planetary Science Conference},
  vol.~41, p. 1383.

\bibitem[{\textit{Ogawa et~al.}(2011)\textit{Ogawa, Matsunaga, Nakamura, Saiki,
  Ohtake, Hiroi, Takeda, Arai, Yokota, Yamamoto, Hirata, Sugihara, Sasaki,
  Haruyama, Morota, Honda, Demura, Kitazato, Terazono, and Asada}}]{Ogawa_2011}
Ogawa, Y., T.~Matsunaga, R.~Nakamura, K.~Saiki, M.~Ohtake, T.~Hiroi, H.~Takeda,
  T.~Arai, Y.~Yokota, S.~Yamamoto, N.~Hirata, T.~Sugihara, S.~Sasaki,
  J.~Haruyama, T.~Morota, C.~Honda, H.~Demura, K.~Kitazato, J.~Terazono, and
  N.~Asada (2011), {The widespread occurrence of high-calcium pyroxene in
  bright-ray craters on the Moon and implications for lunar-crust composition},
  \textit{Geophysical Research Letters}, \textit{38}(17),
  \doi{10.1029/2011GL048569}, l17202.

\bibitem[{\textit{{O'Hara}}(1977)}]{O'Hara_1977}
{O'Hara}, M.~J. (1977), {Geochemical evolution during fractional
  crystallisation of a periodically refilled magma chamber}, \textit{Nature},
  \textit{266}, 503--507, \doi{10.1038/266503a0}.

\bibitem[{\textit{{O'Neill}}(1991)}]{Oneill_1991}
{O'Neill}, H.~S.~C. (1991), {The origin of the moon and the early history of
  the earth - A chemical model. I - The moon. II - The earth},
  \textit{Geochimica et Cosmochimica Acta}, \textit{55}, 1135--1157,
  \doi{10.1016/0016-7037(91)90168-5}.

\bibitem[{\textit{Pahlevan and Stevenson}(2007)}]{Pahlevan_2007}
Pahlevan, K., and D.~J. Stevenson (2007), {Equilibration in the aftermath of
  the lunar-forming giant impact}, \textit{Earth and Planetary Science
  Letters}, \textit{262}, 438--449, \doi{10.1016/j.epsl.2007.07.055}.

\bibitem[{\textit{Pepin}(1991)}]{Pepin_1991}
Pepin, R.~O. (1991), {On the origin and early evolution of terrestrial planet
  atmospheres and meteoritic volatiles}, \textit{Icarus}, \textit{92}(1),
  2--79, \doi{10.1016/0019-1035(91)90036-S}.

\bibitem[{\textit{Philpotts and Schnetzler}(1970)}]{Philpotts_1970}
Philpotts, J., and C.~Schnetzler (1970), {Apollo 11 lunar samples: K, Rb, Sr,
  Ba and rare-earth concentrations in some rocks and separated phases},
  \textit{Geochimica et Cosmochimica Acta Supplement}, \textit{1}, 1471.

\bibitem[{\textit{Piskorz et~al.}(2014)\textit{Piskorz, Elkins-Tanton, and
  Smrekar}}]{Piskorz_2014}
Piskorz, D., L.~T. Elkins-Tanton, and S.~E. Smrekar (2014), {Coronae formation
  on Venus via extension and lithospheric instability}, \textit{Journal of
  Geophysical Research: Planets}, \textit{119}(12), 2568--2582,
  \doi{10.1002/2014JE004636}.

\bibitem[{\textit{Pritchard and Stevenson}(2000)}]{Pritchard_2000}
Pritchard, M., and D.~Stevenson (2000), {Thermal aspects of a lunar origin by
  giant impact}, \textit{Origin of the Earth and Moon}, \textit{1}, 179--196.

\bibitem[{\textit{Rankenburg et~al.}(2006)\textit{Rankenburg, Brandon, and
  Neal}}]{Rankenburg_2006}
Rankenburg, K., A.~D. Brandon, and C.~R. Neal (2006), {Neodymium Isotope
  Evidence for a Chondritic Composition of the Moon}, \textit{Science},
  \textit{312}(5778), 1369--1372, \doi{10.1126/science.1126114}.

\bibitem[{\textit{Ransford and Kaula}(1980)}]{Ransford_1980}
Ransford, G.~A., and W.~M. Kaula (1980), {Heating of the Moon by heterogeneous
  accretion}, \textit{Journal of Geophysical Research: Solid Earth},
  \textit{85}(B11), 6615--6627, \doi{10.1029/JB085iB11p06615}.

\bibitem[{\textit{Rathbun et~al.}(2002)\textit{Rathbun, Spencer, Davies,
  Howell, and Wilson}}]{Rathbun_2002}
Rathbun, J.~A., J.~R. Spencer, A.~G. Davies, R.~R. Howell, and L.~Wilson
  (2002), {Loki, Io: A periodic volcano}, \textit{Geophysical Research
  Letters}, \textit{29}(10), 84--1--84--4, \doi{10.1029/2002GL014747}.

\bibitem[{\textit{Repetto and Nelemans}(2014)}]{Repetto_2014}
Repetto, S., and G.~Nelemans (2014), {The coupled effect of tides and stellar
  winds on the evolution of compact binaries}, \textit{Monthly Notices of the
  Royal Astronomical Society}, \textit{444}(1), 542--557,
  \doi{10.1093/mnras/stu1454}.

\bibitem[{\textit{Reufer et~al.}(2012)\textit{Reufer, Meier, Benz, and
  Wieler}}]{Reufer_2012}
Reufer, A., M.~M. Meier, W.~Benz, and R.~Wieler (2012), {A hit-and-run giant
  impact scenario}, \textit{Icarus}, \textit{221}(1), 296--299,
  \doi{10.1016/j.icarus.2012.07.021}.

\bibitem[{\textit{Riner et~al.}(2009)\textit{Riner, Lucey, Desch, and
  McCubbin}}]{Riner_2009}
Riner, M.~A., P.~G. Lucey, S.~J. Desch, and F.~M. McCubbin (2009), {Nature of
  opaque components on Mercury: Insights into a Mercurian magma ocean},
  \textit{Geophysical Research Letters}, \textit{36}(2),
  \doi{10.1029/2008GL036128}, l02201.

\bibitem[{\textit{Rolf et~al.}(2017)\textit{Rolf, Zhu, W{\"u}nnemann, and
  Werner}}]{Rolf_2017}
Rolf, T., M.-H. Zhu, K.~W{\"u}nnemann, and S.~Werner (2017), {The role of
  impact bombardment history in lunar evolution}, \textit{Icarus},
  \textit{286}, 138 -- 152, \doi{10.1016/j.icarus.2016.10.007}.

\bibitem[{\textit{{Rufu} et~al.}(2017)\textit{{Rufu}, {Aharonson}, and
  {Perets}}}]{Rufu_2017}
{Rufu}, R., O.~{Aharonson}, and H.~B. {Perets} (2017), {A multiple-impact
  origin for the Moon}, \textit{Nature Geoscience}, \textit{10}, 89--94,
  \doi{10.1038/ngeo2866}.

\bibitem[{\textit{Ryder}(2002)}]{Ryder_2002}
Ryder, G. (2002), {Mass flux in the ancient Earth-Moon system and benign
  implications for the origin of life on Earth}, \textit{Journal of Geophysical
  Research: Planets}, \textit{107}(E4), 6--1--6--13,
  \doi{10.1029/2001JE001583}.

\bibitem[{\textit{{Saal} et~al.}(2008)\textit{{Saal}, {Hauri}, {Cascio}, {van
  Orman}, {Rutherford}, and {Cooper}}}]{Saal_2008}
{Saal}, A.~E., E.~H. {Hauri}, M.~L. {Cascio}, J.~A. {van Orman}, M.~C.
  {Rutherford}, and R.~F. {Cooper} (2008), {Volatile content of lunar volcanic
  glasses and the presence of water in the Moon's interior}, \textit{Nature},
  \textit{454}, 192--195, \doi{10.1038/nature07047}.

\bibitem[{\textit{Salmon and Canup}(2012)}]{Salmon_2012}
Salmon, J., and R.~M. Canup (2012), {Lunar Accretion from a Roche-interior
  Fluid Disk}, \textit{The Astrophysical Journal}, \textit{760}(1), 83.

\bibitem[{\textit{Senshu et~al.}(2002)\textit{Senshu, Kuramoto, and
  Matsui}}]{Senshu_2002}
Senshu, H., K.~Kuramoto, and T.~Matsui (2002), {Thermal evolution of a growing
  Mars}, \textit{Journal of Geophysical Research: Planets}, \textit{107}(E12),
  1--1--1--13, \doi{10.1029/2001JE001819}, 5118.

\bibitem[{\textit{Sharp et~al.}(2010)\textit{Sharp, Shearer, McKeegan, Barnes,
  and Wang}}]{Sharp_2010}
Sharp, Z.~D., C.~K. Shearer, K.~D. McKeegan, J.~D. Barnes, and Y.~Q. Wang
  (2010), {The Chlorine Isotope Composition of the Moon and Implications for an
  Anhydrous Mantle}, \textit{Science}, \textit{329}(5995), 1050--1053,
  \doi{10.1126/science.1192606}.

\bibitem[{\textit{{Shirley}}(1983)}]{Shirley_1983}
{Shirley}, D.~N. (1983), {A partially molten magma ocean model}, in
  \textit{Lunar and Planetary Science Conference Proceedings}, vol.~13, edited
  by W.~V. {Boynton} and T.~J. {Ahrens}, pp. A519--A527.

\bibitem[{\textit{{Simonds} et~al.}(1974)\textit{{Simonds}, {Phinney}, and
  {Warner}}}]{Simonds_1974}
{Simonds}, C.~H., W.~C. {Phinney}, and J.~L. {Warner} (1974), {Petrography and
  classification of Apollo 17 non-mare rocks with emphasis on samples from the
  Station 6 boulder}, in \textit{Lunar and Planetary Science Conference
  Proceedings}, vol.~5, pp. 337--353.

\bibitem[{\textit{Snyder et~al.}(1992)\textit{Snyder, Taylor, and
  Neal}}]{Snyder_1992}
Snyder, G.~A., L.~A. Taylor, and C.~R. Neal (1992), {A chemical model for
  generating the sources of mare basalts: Combined equilibrium and fractional
  crystallization of the lunar magmasphere}, \textit{Geochimica et Cosmochimica
  Acta}, \textit{56}(10), 3809 -- 3823, \doi{10.1016/0016-7037(92)90172-F}.

\bibitem[{\textit{Solomon}(1980)}]{Solomon_1980}
Solomon, S.~C. (1980), {Differentiation of crusts and cores of the terrestrial
  planets: Lessons for the early Earth?}, \textit{Precambrian Research},
  \textit{10}(3--4), 177--194, \doi{10.1016/0301-9268(80)90011-X}, comparative
  Planetary Evolution: Implications for the Proto-Archean.

\bibitem[{\textit{Solomon}(1986)}]{Solomon_1986}
Solomon, S.~C. (1986), {On the early thermal state of the Moon}, in
  \textit{Origin of the Moon}, pp. 435--452.

\bibitem[{\textit{{Solomon} and {Chaiken}}(1976)}]{Solomon_1976}
{Solomon}, S.~C., and J.~{Chaiken} (1976), {Thermal expansion and thermal
  stress in the moon and terrestrial planets--Clues to early thermal history},
  in \textit{Lunar and Planetary Science Conference Proceedings}, vol.~7,
  edited by D.~C. {Kinsler}, pp. 3229--3243.

\bibitem[{\textit{{Solomon} and {Longhi}}(1977)}]{Solomon_1977b}
{Solomon}, S.~C., and J.~{Longhi} (1977), {Magma oceanography: 1.--Thermal
  evolution}, in \textit{Lunar and Planetary Science Conference Proceedings},
  vol.~8, edited by R.~B. {Merril}, pp. 583--599.

\bibitem[{\textit{Spera}(1992)}]{Spera_1992}
Spera, F.~J. (1992), {Lunar magma transport phenomena}, \textit{Geochimica et
  Cosmochimica Acta}, \textit{56}(6), 2253 -- 2265,
  \doi{10.1016/0016-7037(92)90187-N}.

\bibitem[{\textit{Spicuzza et~al.}(2007)\textit{Spicuzza, Day, Taylor, and
  Valley}}]{Spicuzza_2007}
Spicuzza, M.~J., J.~M. Day, L.~A. Taylor, and J.~W. Valley (2007), {Oxygen
  isotope constraints on the origin and differentiation of the Moon},
  \textit{Earth and Planetary Science Letters}, \textit{253}(1--2), 254--265,
  \doi{10.1016/j.epsl.2006.10.030}.

\bibitem[{\textit{Squyres et~al.}(1988)\textit{Squyres, Reynolds, Summers, and
  Shung}}]{Squyres_1988}
Squyres, S.~W., R.~T. Reynolds, A.~L. Summers, and F.~Shung (1988),
  {Accretional heating of the satellites of Saturn and Uranus}, \textit{Journal
  of Geophysical Research: Solid Earth}, \textit{93}(B8), 8779--8794,
  \doi{10.1029/JB093iB08p08779}.

\bibitem[{\textit{Stevenson et~al.}(1986)\textit{Stevenson, Harris, and
  Lunine}}]{Stevenson_1986}
Stevenson, D., A.~Harris, and J.~Lunine (1986), {Origins of satellites}, in
  \textit{IAU Colloq. 77: Some Background about Satellites}, pp. 39--88.

\bibitem[{\textit{Takeda and Ida}(2001)}]{Takeda_2001}
Takeda, T., and S.~Ida (2001), {Angular Momentum Transfer in a Protolunar
  Disk}, \textit{The Astrophysical Journal}, \textit{560}(1), 514.

\bibitem[{\textit{{Tanaka} et~al.}(1996)\textit{{Tanaka}, {Inaba}, and
  {Nakazawa}}}]{Tanaka_1996}
{Tanaka}, H., S.~{Inaba}, and K.~{Nakazawa} (1996), {Steady-State Size
  Distribution for the Self-Similar Collision Cascade}, \textit{Icarus},
  \textit{123}, 450--455, \doi{10.1006/icar.1996.0170}.

\bibitem[{\textit{Taylor and Wieczorek}(2014)}]{Taylor_2014}
Taylor, G.~J., and M.~A. Wieczorek (2014), {Lunar bulk chemical composition: a
  post-Gravity Recovery and Interior Laboratory reassessment},
  \textit{Philosophical Transactions of the Royal Society of London A:
  Mathematical, Physical and Engineering Sciences}, \textit{372}(2024),
  \doi{10.1098/rsta.2013.0242}.

\bibitem[{\textit{{Taylor}}(2014)}]{Taylor_2014b}
{Taylor}, S.~R. (2014), {The Moon re-examined}, \textit{Geochimica et
  Cosmochimica Acta}, \textit{141}, 670--676, \doi{10.1016/j.gca.2014.06.031}.

\bibitem[{\textit{{Taylor} et~al.}(1993)\textit{{Taylor}, {Norman}, and
  {Esat}}}]{Taylor_1993}
{Taylor}, S.~R., M.~D. {Norman}, and T.~M. {Esat} (1993), {The Mg-suite and the
  highland crust: an unsolved enigma}, in \textit{Lunar and Planetary Science
  Conference}, vol.~24.

\bibitem[{\textit{Taylor et~al.}(2006{\natexlab{a}})\textit{Taylor, Taylor, and
  Taylor}}]{Taylor_2006}
Taylor, S.~R., G.~J. Taylor, and L.~A. Taylor (2006{\natexlab{a}}), {The Moon:
  A Taylor perspective}, \textit{Geochimica et Cosmochimica Acta},
  \textit{70}(24), 5904 -- 5918, \doi{10.1016/j.gca.2006.06.262}, a Special
  Issue Dedicated to Larry A. Haskin.

\bibitem[{\textit{Taylor et~al.}(2006{\natexlab{b}})\textit{Taylor, Pieters,
  and MacPherson}}]{Taylor_2006b}
Taylor, S.~R., C.~M. Pieters, and G.~J. MacPherson (2006{\natexlab{b}}),
  {Earth-Moon System, Planetary Science, and Lessons Learned}, \textit{Reviews
  in Mineralogy and Geochemistry}, \textit{60}(1), 657,
  \doi{10.2138/rmg.2006.60.7}.

\bibitem[{\textit{Tian et~al.}(2017)\textit{Tian, Wisdom, and
  Elkins-Tanton}}]{Tian_2017}
Tian, Z., J.~Wisdom, and L.~Elkins-Tanton (2017), {Coupled orbital-thermal
  evolution of the early Earth-Moon system with a fast-spinning Earth},
  \textit{Icarus}, \textit{281}, 90 -- 102, \doi{10.1016/j.icarus.2016.08.030}.

\bibitem[{\textit{{Touboul} et~al.}(2007)\textit{{Touboul}, {Kleine},
  {Bourdon}, {Palme}, and {Wieler}}}]{Touboul_2007}
{Touboul}, M., T.~{Kleine}, B.~{Bourdon}, H.~{Palme}, and R.~{Wieler} (2007),
  {Late formation and prolonged differentiation of the Moon inferred from W
  isotopes in lunar metals}, \textit{Nature}, \textit{450}, 1206--1209,
  \doi{10.1038/nature06428}.

\bibitem[{\textit{Touboul et~al.}(2009)\textit{Touboul, Kleine, Bourdon, Palme,
  and Wieler}}]{Touboul_2009}
Touboul, M., T.~Kleine, B.~Bourdon, H.~Palme, and R.~Wieler (2009), {Tungsten
  isotopes in ferroan anorthosites: Implications for the age of the Moon and
  lifetime of its magma ocean}, \textit{Icarus}, \textit{199}(2), 245--249,
  \doi{10.1016/j.icarus.2008.11.018}.

\bibitem[{\textit{{Touma} and {Wisdom}}(1994)}]{Touma_1994}
{Touma}, J., and J.~{Wisdom} (1994), {Evolution of the Earth-Moon system},
  \textit{Astron. J.}, \textit{108}, 1943--1961, \doi{10.1086/117209}.

\bibitem[{\textit{Touma and Wisdom}(1998)}]{Touma_1998}
Touma, J., and J.~Wisdom (1998), {Resonances in the Early Evolution of the
  Earth-Moon System}, \textit{The Astronomical Journal}, \textit{115}(4), 1653.

\bibitem[{\textit{{Turcotte} and {Schubert}}(2014)}]{Turcotte_2014}
{Turcotte}, D.~L., and G.~{Schubert} (2014), \textit{{Geodynamics - 3rd
  Edition}}, 636 pp., Cambridge University Press.

\bibitem[{\textit{Vander~Kaaden and McCubbin}(2015)}]{VanderKaaden_2015}
Vander~Kaaden, K.~E., and F.~M. McCubbin (2015), {Exotic crust formation on
  Mercury: Consequences of a shallow, FeO-poor mantle}, \textit{Journal of
  Geophysical Research: Planets}, \textit{120}(2), 195--209,
  \doi{10.1002/2014JE004733}.

\bibitem[{\textit{Wakita and Schmitt}(1970)}]{Wakita_1970}
Wakita, H., and R.~A. Schmitt (1970), {Lunar Anorthosites: Rare-Earth and Other
  Elemental Abundances}, \textit{Science}, \textit{170}(3961), 969--974,
  \doi{10.1126/science.170.3961.969}.

\bibitem[{\textit{Walker}(1983)}]{Walker_1983}
Walker, D. (1983), {Lunar and terrestrial crust formation}, \textit{Journal of
  Geophysical Research: Solid Earth}, \textit{88}(S01), B17--B25,
  \doi{10.1029/JB088iS01p00B17}.

\bibitem[{\textit{{Walker} and {Kiefer}}(1985)}]{Walker_1985}
{Walker}, D., and W.~S. {Kiefer} (1985), {Xenolith digestion in large magma
  bodies}, in \textit{Lunar and Planetary Science Conference Proceedings},
  vol.~15, edited by G.~{Ryder} and G.~{Schubert}, pp. C585--C590.

\bibitem[{\textit{{Walker} et~al.}(1975)\textit{{Walker}, {Longhi}, and
  {Hays}}}]{Walker_1975}
{Walker}, D., J.~{Longhi}, and J.~F. {Hays} (1975), {Differentiation of a very
  thick magma body and implications for the source regions of mare basalts}, in
  \textit{Lunar and Planetary Science Conference Proceedings}, vol.~6, pp.
  1103--1120.

\bibitem[{\textit{{Walker} et~al.}(1980)\textit{{Walker}, {Hager}, and
  {Hayes}}}]{Walker_1980}
{Walker}, D., B.~H. {Hager}, and J.~F. {Hayes} (1980), {Mass and Heat Transport
  in a Lunar Magma Ocean by Sinking Blobs}, in \textit{Lunar and Planetary
  Inst.~Technical Report}, vol.~11, pp. 1196--1198.

\bibitem[{\textit{{Warner} et~al.}(1977)\textit{{Warner}, {Phinney}, {Bickel},
  and {Simonds}}}]{Warner_1977}
{Warner}, J.~L., W.~C. {Phinney}, C.~E. {Bickel}, and C.~H. {Simonds} (1977),
  {Feldspathic granulitic impactites and pre-final bombardment lunar
  evolution}, in \textit{Lunar and Planetary Science Conference Proceedings},
  vol.~8, edited by R.~B. {Merril}, pp. 2051--2066.

\bibitem[{\textit{Warren}(1985)}]{Warren_1985}
Warren, P.~H. (1985), {The Magma Ocean Concept and Lunar Evolution},
  \textit{Annual Review of Earth and Planetary Sciences}, \textit{13}(1),
  201--240, \doi{10.1146/annurev.ea.13.050185.001221}.

\bibitem[{\textit{Warren and Wasson}(1979)}]{Warren_1979}
Warren, P.~H., and J.~T. Wasson (1979), {The origin of KREEP}, \textit{Reviews
  of Geophysics}, \textit{17}(1), 73--88, \doi{10.1029/RG017i001p00073}.

\bibitem[{\textit{Wieczorek et~al.}(2013)\textit{Wieczorek, Neumann, Nimmo,
  Kiefer, Taylor, Melosh, Phillips, Solomon, Andrews-Hanna, Asmar, Konopliv,
  Lemoine, Smith, Watkins, Williams, and Zuber}}]{Wieczorek_2013}
Wieczorek, M.~A., G.~A. Neumann, F.~Nimmo, W.~S. Kiefer, G.~J. Taylor, H.~J.
  Melosh, R.~J. Phillips, S.~C. Solomon, J.~C. Andrews-Hanna, S.~W. Asmar,
  A.~S. Konopliv, F.~G. Lemoine, D.~E. Smith, M.~M. Watkins, J.~G. Williams,
  and M.~T. Zuber (2013), {The Crust of the Moon as Seen by GRAIL},
  \textit{Science}, \textit{339}(6120), 671--675,
  \doi{10.1126/science.1231530}.

\bibitem[{\textit{Williams et~al.}(2014)\textit{Williams, Konopliv, Boggs,
  Park, Yuan, Lemoine, Goossens, Mazarico, Nimmo, Weber, Asmar, Melosh,
  Neumann, Phillips, Smith, Solomon, Watkins, Wieczorek, Andrews-Hanna, Head,
  Kiefer, Matsuyama, McGovern, Taylor, and Zuber}}]{Williams_2014}
Williams, J.~G., A.~S. Konopliv, D.~H. Boggs, R.~S. Park, D.-N. Yuan, F.~G.
  Lemoine, S.~Goossens, E.~Mazarico, F.~Nimmo, R.~C. Weber, S.~W. Asmar, H.~J.
  Melosh, G.~A. Neumann, R.~J. Phillips, D.~E. Smith, S.~C. Solomon, M.~M.
  Watkins, M.~A. Wieczorek, J.~C. Andrews-Hanna, J.~W. Head, W.~S. Kiefer,
  I.~Matsuyama, P.~J. McGovern, G.~J. Taylor, and M.~T. Zuber (2014), {Lunar
  interior properties from the GRAIL mission}, \textit{Journal of Geophysical
  Research: Planets}, \textit{119}(7), 1546--1578, \doi{10.1002/2013JE004559},
  2013JE004559.

\bibitem[{\textit{Wolf and Anders}(1980)}]{Wolf_1980}
Wolf, R., and E.~Anders (1980), {Moon and Earth : compositional differences
  inferred from siderophiles, volatiles, and alkalis in basalts},
  \textit{Geochimica et Cosmochimica Acta}, \textit{44}(12), 2111--2124,
  \doi{10.1016/0016-7037(80)90208-2}.

\bibitem[{\textit{{Wood} et~al.}(1970{\natexlab{a}})\textit{{Wood}, {Dickey},
  {Marvin}, and {Powell}}}]{Wood_1970a}
{Wood}, J.~A., J.~S. {Dickey}, Jr., U.~B. {Marvin}, and B.~N. {Powell}
  (1970{\natexlab{a}}), {Lunar Anorthosites}, \textit{Science},
  \textit{167}(3918), 602--604, \doi{10.1126/science.167.3918.602}.

\bibitem[{\textit{{Wood} et~al.}(1970{\natexlab{b}})\textit{{Wood}, {Dickey},
  {Marvin}, and {Powell}}}]{Wood_1970b}
{Wood}, J.~A., J.~S. {Dickey}, Jr., U.~B. {Marvin}, and B.~N. {Powell}
  (1970{\natexlab{b}}), {Lunar anorthosites and a geophysical model of the
  moon}, \textit{Geochimica et Cosmochimica Acta Supplement}, \textit{1}, 965.

\bibitem[{\textit{{Wyatt} et~al.}(2011)\textit{{Wyatt}, {Clarke}, and
  {Booth}}}]{Wyatt_2011}
{Wyatt}, M.~C., C.~J. {Clarke}, and M.~{Booth} (2011), {Debris disk size
  distributions: steady state collisional evolution with Poynting-Robertson
  drag and other loss processes}, \textit{Celestial Mechanics and Dynamical
  Astronomy}, \textit{111}, 1--28, \doi{10.1007/s10569-011-9345-3}.

\bibitem[{\textit{Yamamoto et~al.}(2012)\textit{Yamamoto, Nakamura, Matsunaga,
  Ogawa, Ishihara, Morota, Hirata, Ohtake, Hiroi, Yokota, and
  Haruyama}}]{Yamamoto_2012}
Yamamoto, S., R.~Nakamura, T.~Matsunaga, Y.~Ogawa, Y.~Ishihara, T.~Morota,
  N.~Hirata, M.~Ohtake, T.~Hiroi, Y.~Yokota, and J.~Haruyama (2012), {Massive
  layer of pure anorthosite on the Moon}, \textit{Geophysical Research
  Letters}, \textit{39}(13), \doi{10.1029/2012GL052098}, l13201.

\bibitem[{\textit{Yamamoto et~al.}(2015)\textit{Yamamoto, Nakamura, Matsunaga,
  Ogawa, Ishihara, Morota, Hirata, Ohtake, Hiroi, Yokota, and
  Haruyama}}]{Yamamoto_2015b}
Yamamoto, S., R.~Nakamura, T.~Matsunaga, Y.~Ogawa, Y.~Ishihara, T.~Morota,
  N.~Hirata, M.~Ohtake, T.~Hiroi, Y.~Yokota, and J.~Haruyama (2015), {Global
  occurrence trend of high-Ca pyroxene on lunar highlands and its
  implications}, \textit{Journal of Geophysical Research: Planets},
  \textit{120}(5), 831--848, \doi{10.1002/2014JE004740}.

\bibitem[{\textit{{Zhang} et~al.}(2012)\textit{{Zhang}, {Dauphas}, {Davis},
  {Leya}, and {Fedkin}}}]{Zhang_2012}
{Zhang}, J., N.~{Dauphas}, A.~M. {Davis}, I.~{Leya}, and A.~{Fedkin} (2012),
  {The proto-Earth as a significant source of lunar material}, \textit{Nature
  Geoscience}, \textit{5}, 251--255, \doi{10.1038/ngeo1429}.

\bibitem[{\textit{Zhang et~al.}(2013)\textit{Zhang, Parmentier, and
  Liang}}]{Zhang_2013}
Zhang, N., E.~M. Parmentier, and Y.~Liang (2013), {A 3-D numerical study of the
  thermal evolution of the Moon after cumulate mantle overturn: The importance
  of rheology and core solidification}, \textit{Journal of Geophysical
  Research: Planets}, \textit{118}(9), 1789--1804, \doi{10.1002/jgre.20121}.

\end{thebibliography}

\appendix
\section{Calculating the LMO Convective Flux}
\label{Apd:Nusselt}

For this work we used the Nusselt number (\textit{Nu}) to calculate the convective flux of the LMO (Equation \ref{eq:Nusselt}). \citet{Breuer_2015} suggest that depending on the convecting layer's geometry, mode of heating, and boundary conditions, \textit{a} may range from 0.195 to 0.339 and $\beta$ may range from 1/4 to 1/3. They also note that Equation \ref{eq:Nusselt} is only valid if the change in viscosity in the convecting layer is small. We use \textit{a} = 0.124 and $\beta$ = 0.309 from experimental work by \citet{Niemela_2000}. It is worth noting that \textit{Ra} is typically $\sim$~$10^{22}$ for the LMO, which is considerably higher that what is achievable by experiments. The highest \textit{Ra} experiments are also conducted with liquid helium \citep{Niemela_2000}, which is a rather different environment to the LMO. There is an alternative method of calculating the convective heat flux. \citet{Neumann_2014} and \citet{Monteux_2016} used an effective thermal conductivity for convection so that they could use the Fourier law formulation. \textcolor{black}{It is not clear whether this would lead to a more accurate estimate however and so} we use the \textit{Nu} procedure stated above for this work.

\section{\textcolor{black}{Convergence Tests}}
\label{Apd:Convergence}

\begin{figure}[ht]
\includegraphics[width=0.996\textwidth]{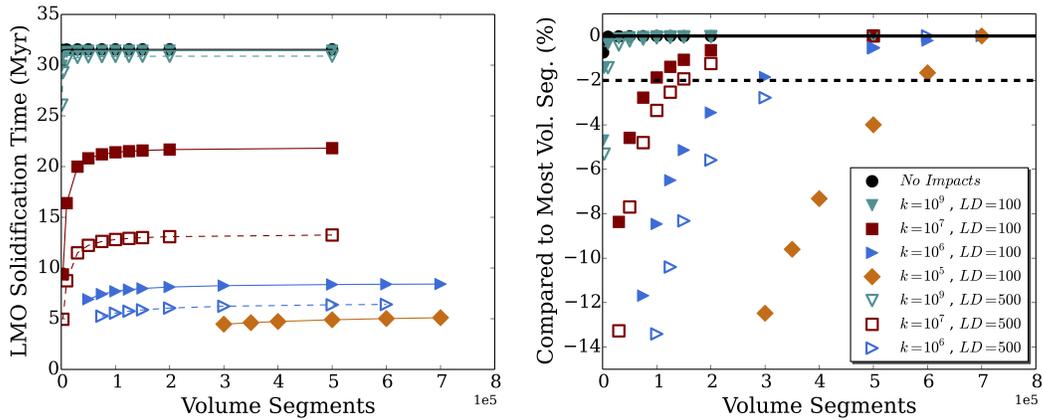}
\centering
\caption{Convergence test of the LMO solidification time (on the left) for varying levels of impact intensity (i.e. $k$) as a function of the number of volume segments. Shown on right is the difference between a solidification time and the solidification time with the most volume segments for a particular set. Colored markers and lines are used for the different $k$ values. The filled markers and solid lines correspond to the case when the largest debris (LD) is 100~km. The open markers and dashed lines correspond to the case when the LD is 500~km. Black filled circles show the no impacts runs. Note that these are almost completely overlain by the $k$~=~$10^9$~kg/m\textsuperscript{2} points. The solid black line marks the 0\% point and the dashed black line marks the -2\% point (the point at which a particular $k$ value is considered converged).}
\label{fig:Convergence}
\end{figure}

It is important to ensure that a sufficient number of volume segments (and equivalently sufficiently small timesteps) is used such that our results are not dependent on the number of volume segments used. In particular, since the mass added to the Moon during any given numerical step is dependent on the timestep, too few volume segments (equivalently timesteps that are too large) will generate very large holes. The conservation and redistribution of crustal material will usually ensure that saturation, i.e. the total area of holes approaching the surface area of the Moon, is appropriately handled.  If, however, the area of holes added in any time step is a large fraction of the surface area of the Moon there is a danger that this will break down and cause the total area of holes to exceed the surface area of the Moon.  Since the conversion of impacting mass into hole area is governed by $k$ the number of volume segments required for convergence will also depend on $k$. Therefore, we conducted a number of convergence tests by varying the impact intensity (i.e. $k$) and the number of volume segments to find the minimum number of volume segments required. For all convergence tests we used our nominal values listed in Table \ref{tableConstants}. Figure \ref{fig:Convergence} shows the LMO solidification time for our convergence tests. Overall, the less intense the bombardment (i.e. higher $k$ values which produce smaller hole areas) the fewer volume segments that are needed for convergence, as expected. When the largest debris is 100~km, for $k \geq 10^7$~kg/m\textsuperscript{2}, we find that $10^{5}$ volume segments is sufficient. For $k=10^6$~kg/m\textsuperscript{2} we find that $3 \times 10^{5}$ volume segments is sufficient and for $k=10^5$~kg/m\textsuperscript{2} we find that $6 \times 10^{5}$ volume segments is sufficient. When the largest debris is 500~km, for $k \geq 10^7$~kg/m\textsuperscript{2} we find that $1.5 \times 10^{5}$ volume segments is sufficient and for $k=10^6$~kg/m\textsuperscript{2} we find that $5 \times 10^{5}$ volume segments is sufficient.

\section{\textcolor{black}{Model Parameter Sensitivity}}
\label{Apd:Sensitivity}

\begin{figure}
\includegraphics[width=1.10\textwidth]{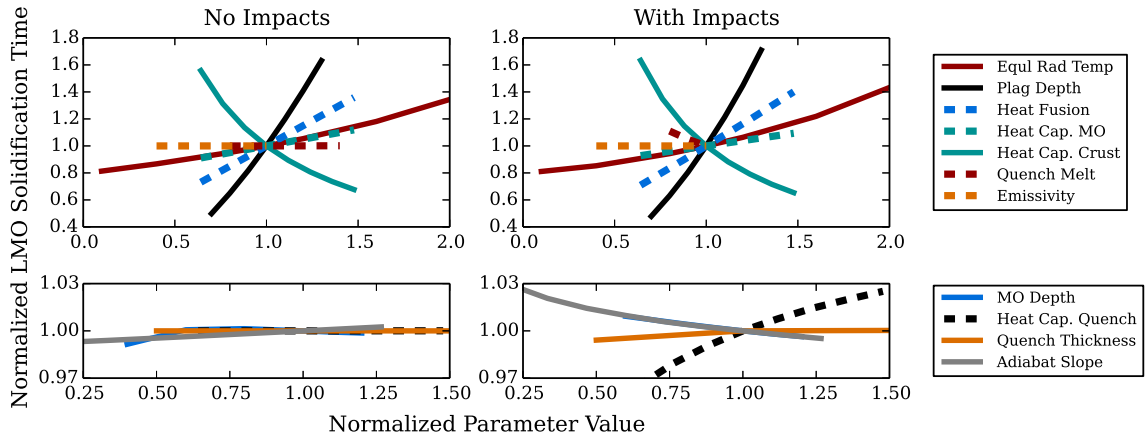}
\centering
\caption{\textcolor{black}{Sensitivity tests for the change of the LMO solidification time from the nominal value due to the variation of one input parameter at a time. The no impacts runs are shown on the left and the with impacts runs ($k$~=~$10^{7}$~kg/m\textsuperscript{2}) are shown on the right. For clarity, the figure is split so that the larger changes in LMO solidification time are shown on the top panels and the smaller changes in LMO solidification time are shown on the bottom panels. The nominal parameter values are listed on Table \ref{tableConstants} and the nominal LMO solidification times are 31.6~Myr for the no impacts case and 21.4~Myr for the impacts case.}}
\label{fig:ParameterSensitivity}
\end{figure}

Although we use the nominal input parameter values listed in Table \ref{tableConstants} for the majority of our work, some of those parameters are subject to uncertainty. For example, estimates for the percentage by volume of the LMO that has to be solidified prior to plagioclase stability vary from 60\% to 80\% \citep{Longhi_1980, Snyder_1992, ElkinsTanton_2011, Lin_2017b}. Thus, it is important to evaluate the sensitivity of output parameters, particularly the solidification time of the LMO, to variability of input parameters. In Figure \ref{fig:ParameterSensitivity} we show the change in LMO solidification time as a function of varying some of the model input parameters (with the exception of $k$ and the LMO dynamic viscosity). Dynamic viscosity is not shown since it was varied over eight orders of magnitude.

Some input parameters have the same effect on the LMO solidification time whether impacts are or are not included. Three parameters, the maximum thickness of quench crust, the initial depth of the LMO, and emissivity have (nearly) no effect on the solidification time in either case. The insensitivity of the solidification time to the initial LMO depth is consistent with \citet{Solomon_1977b}. This implies that having precise values for these three parameters is not vital. The equilibrium radiative temperature, depth at which plagioclase formation begins, and the heat of fusion and heat capacity of the LMO all correlate positively with the solidification time (i.e. increasing them increases the solidification time), which is as expected. Increasing the equilibrium radiative temperature, and increasing the depth at which plagioclase formation begins (which increases the final depth of the crust) both act to slow down the release of thermal energy from the LMO. Increasing the heat of fusion or the heat capacity of the LMO increases the total thermal energy that must be released during the solidification process. On the other hand, the heat capacity of the crust is negatively correlated with the solidification time. A higher heat capacity in the crust increases the conductive flux through the crust, and so we would expect it to decrease the solidification time.

Other parameters have different effects on the LMO solidification time depending on if there are impact generated holes or not. Dynamic viscosity and the melting temperature of quench crust have no effect on the solidification time when there are no impacts; however, when there are impacts dynamic viscosity is positively correlated and the melting temperature is negatively correlated with the solidification time. When there are impacts, varying dynamic viscosity from 1 to $10^8$~Pa$\cdot$s results in a -35\% to 7\% change in the solidification time. Lower dynamic viscosity values would increase \textit{Ra}, which through \textit{Nu} would lead to a thinner quench layer (see Section \ref{sec:Methods:Code:Quench}) and thus would decrease the solidification time by increasing the thermal flux. Lower values of the melting temperature of quench crust will have the opposite effect since it will reduce the conductive flux through quench crust by decreasing the temperature at the bottom of quench crust (i.e. its melting temperature). The heat capacity of the quench crust and the slope of the adiabat also have no effect on the solidification time when there are no impacts but when there are impacts, they have a small correlation with the solidification time. The positive correlation of the quench heat capacity arises for exactly the same reason as the positive correlation with the heat capacity of the plagioclase floatation crust. On the other hand, the negative correlation of the adiabat slope is due to smaller values reducing the temperature difference between the bottom and top of the LMO and thus decreasing the thermal flux out of the LMO.

From Figure \ref{fig:ParameterSensitivity} we can see that the LMO solidification time is most sensitive to the depth at which plagioclase starts to form and the heat capacity of the crust. This is consistent with the work of \citet{Minear_1978}. As mentioned previously, there is a range of estimated LMO depths at which plagioclase becomes stable. The depth could be about our nominal value (i.e. 100~km) or as deep as 250~km \citep{Longhi_1980, Snyder_1992, ElkinsTanton_2011, Lin_2017b}. Since we do not model the geochemistry, we limit our variable change of the plagioclase stability depth since varying it significantly changes the final crustal thickness. With or without impacts, when we set the depth to 70~km, the final crustal thickness was 31~km and when we set the depth to 130~km, the final crustal thickness was 58~km. It is plausible that the thickness of the primordial lunar crust was greater than the average crustal thickness today. However, we do not explore that possibility in this work. Overall, our results indicate that the LMO solidification is primarily governed by the conductive flux through the crust, which is both a function of its thickness and its thermal properties (i.e. specific heat capacity).

%%%
\listofchanges
%%%

\end{document}